\documentclass[fleqn,usenatbib]{mnras}

\usepackage{mathptmx}
\usepackage[T1]{fontenc}
\usepackage{graphicx}	% Including figure files
\usepackage{amsmath}	% Advanced maths commands
\usepackage{amssymb}	% Extra maths symbols
\usepackage[usenames,dvipsnames]{xcolor} %remove after revision
\usepackage[normalem]{ulem} % To strike out text
\usepackage{hyperref,siunitx}
\newcommand{\eagle}{\mbox{\sc{Eagle}}}

\newcommand{\jwst}{\mbox{\it JWST}}
\newcommand{\flares}{\mbox{\sc Flares}}

\newcommand{\Halpha}{\mbox{$\mathrm{H}_{\alpha}$}}
\newcommand{\Hbeta}{\mbox{$\mathrm{H}_{\beta}$}}
\newcommand{\Hgamma}{\mbox{$\mathrm{H}_{\gamma}$}}
\newcommand{\um}{\SI{}{\micro\meter}}
\newcommand{\eg}[0]{$\textnormal{e.g. }$}
\newcommand{\ie}[0]{$\textnormal{i.e. }$}
\newcommand{\etc}[0]{$\textnormal{etc}$}
\definecolor{deepchestnut}{rgb}{0.73, 0.31, 0.28}

\defcitealias{Lovell2021}{\flares\,I}
\title[Consequences of star-dust geometry]{First Light And Reionisation Epoch Simulations (FLARES) XII: \newline The consequences of star-dust geometry on galaxies in the EoR}

\author[Aswin P. Vijayan et al.]{Aswin P. Vijayan$^{1,2}$\thanks{E-mail: apavi@space.dtu.dk},
Peter A.~Thomas$^{3}$,
Christopher C. Lovell$^{4,3}$, 
Stephen M. Wilkins$^{3}$,
\newauthor
Thomas R. Greve$^{1,2}$,
Dimitrios Irodotou$^{5}$,
William J. Roper$^{3}$,
Louise T. C. Seeyave$^{3}$
\\
$^{1}$Cosmic Dawn Center (DAWN)\\
$^{2}$DTU-Space, Technical University of Denmark, Elektrovej 327, DK-2800 Kgs. Lyngby, Denmark\\
$^{3}$Astronomy Centre, University of Sussex, Falmer, Brighton BN1 9QH, UK\\
$^{4}$Institute of Cosmology and Gravitation, University of Portsmouth, Burnaby Road, Portsmouth, PO1 3FX, UK\\
$^{5}$Department of Physics, University of Helsinki, Gustaf Hällströmin katu 2, FI-00014, Helsinki, Finland\\
}

\date{Accepted XXX. Received YYY; in original form ZZZ}

\pubyear{2023}
\begin{document}
\label{firstpage}
\pagerange{\pageref{firstpage}--\pageref{lastpage}}
\maketitle

\begin{abstract}
Using the First Light And Reionisation Epoch Simulations (\flares), a suite of hydrodynamical simulations we explore the consequences of a realistic model for star--dust geometry on the observed properties of galaxies. We find that the UV attenuation declines rapidly from the central regions of galaxies, and bright galaxies have spatially extended star formation that suffers less obscuration than their fainter counterparts, demonstrating a non-linear relationship between the UV luminosity and the UV attenuation, giving a double power-law shape to the UVLF. Spatially distinct stellar populations within galaxies experience a wide range of dust attenuation due to variations in the dust optical depth along their line-of-sight; which can range from completely dust obscured to being fully unobscured. The overall attenuation curve of a galaxy is then a complex combination of various lines-of-sight within the galaxy. We explore the manifestation of this effect to study the reliability of line ratios to infer galaxy properties, in particular the Balmer decrement and the BPT diagram. We find the Balmer decrement predicted Balmer line attenuation to be higher (factor of $1$ to $\gtrsim10$) than expected from commonly used attenuation curves.
The observed BPT line ratios deviate from their intrinsic values (median difference of 0.08 (0.02) and standard deviation of 0.2 (0.05) for log$_{10}$([N$\textsc{ii}$]$\lambda 6585/$H$_{\alpha}$) (log$_{10}$([O\textsc{iii}]$\lambda 5008/$H$_{\beta}$)).
Finally, we explore the variation in observed properties (UV attenuation, UV slope and Balmer decrement) with viewing angle, finding average differences of $\sim0.3$ magnitudes in the UV attenuation. 
\end{abstract}


%%%%%%%%%%%%%%%%%%%%%%%%%%%%%%%%%%%%%%%%%%%%%%%%%%

%%%%%%%%%%%%%%%%% BODY OF PAPER %%%%%%%%%%%%%%%%%%

\section{Introduction}\label{sec:intro}
Attenuation by dust is ubiquitous when performing any astronomical observations in the rest-frame ultraviolet (UV) to the near-infrared (NIR), since dust is one of the key components of the interstellar medium (ISM) of galaxies. Dust modifies the intrinsic emission of galaxies, by partly absorbing light from the UV to the NIR, and re-emitting at longer wavelengths. Even though dust only constitutes a small fraction of the baryonic mass of a galaxy, it is estimated that nearly half of the photons in the Universe have been reprocessed by dust \cite[]{Dwek1998,Bernstein2002,Driver2016}. In order to derive the physical properties of galaxies, such as their stellar mass, gas content, star formation rate, age and metallicity, it is important to separate the impact of dust from the intrinsic stellar and nebular emission \cite[see reviews by][]{Walcher2011_review,Conroy2013_review}.
% , the \textit{deus ex machina} for multiple studies. 

In the current observational astronomy paradigm this decoupling is usually achieved through the use of dust attenuation curves/laws\footnote{In this work we will use attenuation laws or attenuation curves interchangeably}, which parameterise the wavelength dependence of the dust optical depth. These curves or laws are a function of the dust grain properties, such as their size, shape and composition, as well as the relative geometry of stars and dust (star-dust geometry) in the galaxy \cite[see review by][]{Salim_Narayanan_2020}. Dust attenuation includes the aggregate effect of dust absorbing stellar and nebular emission along the line-of-sight, as well as the influence of unobstructed starlight that has not undergone any processing. Additionally, it encompasses the scattering of light into and out of the line-of-sight. Various works have measured dust attenuation curves or laws in the local Universe, such as those observed in the Milky Way \cite[MW,][]{MWcurve1990} or the Magellanic Clouds \cite[LMC and SMC,][]{Pei1992}. Generally, these are observed extinction laws, which corresponds to the attenuation law through a distant uniform screen geometry. This geometry is unrealistic for real galaxies; despite this, these curves are still regularly used to correct for dust attenuation of the spectral energy distribution (SED) of distant galaxies.

Real galaxies exhibit complex geometries, where the assumption of a uniform dust screen is insufficient. Thus, there is a concerted effort to obtain average / effective attenuation laws, that encompass the impact of different dust grain properties and compositions (silicates, carbonates, \etc), along with the relative distribution of stars and dust within galaxies. These two aspects are hard to disentangle when dealing with unresolved observations, or when lacking multi-wavelength coverage. One such average dust attenuation law, widely used in the literature, is known as the Calzetti-starburst attenuation curve. This curve has been measured for bright star forming galaxies \cite[]{Calzetti1994,Calzetti1997}, and it is flatter than the SMC dust law. 

Theoretical studies have also focused on geometries other than a simple uniform screen by taking into account the clumpiness of the dust and stellar distribution \cite[\eg][]{Witt1992,Witt2000,Inoue2005,Chevallard2013}. Such a scenario leads to a more transparent medium than a uniform screen, due to stellar light escaping through dust free paths. These studies have also demonstrated that the attenuation law through a clumpy medium with a SMC type dust composition can be very similar to the Calzetti type dust law, which is significantly greyer at shorter wavelengths. Thus, galaxies which have intrinsically different dust properties can exhibit attenuation curves which are similar.

When modelling galaxy SEDs using various fitting methods, attenuation or extinction curves are commonly used to correct for the effect of dust. To simplify the fitting process, these curves are generally used in conjunction with the assumption that stars sit behind a uniform screen of dust (with the choice of the attenuation curve describing the effective attenuation through the screen, or how porous it is) with higher dust optical depths and different grain properties towards young star forming regions \cite[\eg][]{CF00}. Therefore, SED fitting already assumes a particular geometry between stars and dust, as well as dust properties, \textit{a priori} when any particular dust curve is chosen. 

Recent years have seen a significant increase in the number of detected sources in the high-redshift Universe ($z\ge5$). 
This has been mainly due to multi-wavelength observations taken through numerous ground and space based observatories, which has provided significant insight into the UV-optical emission of these galaxies. 
However, there are many obstacles when constructing a comprehensive picture of galaxy formation and evolution in the early Universe. 
This has been primarily due to the limitations of ground based observatories in terms of their accessible wavelength range for observing high-$z$ galaxy spectral features (\eg\ Lyman or Balmer break redshifting out of the observed wavelength, or not having rest-frame UV, optical and near-IR coverage to simultaneously constrain stellar masses and star formation rates).
In addition, space based observatories are, generally, relatively smaller compared to their ground based counterparts, leading to poor spatial resolution as well as sensitivity. 

As discussed, another obstacle is the presence of dust in galaxies. 
Even though the dust content of high-redshift galaxies is expected to be smaller than in the local Universe or at cosmic noon ($z \sim 2$), recent studies of the high-redshift Universe have revealed a number of sources with highly dust-obscured star formation \cite[\eg][]{Fudamoto2021,Ferrara2022,Algera2023}. 
Understanding the physical properties of high-$z$ galaxies is also complicated by the lack of spatially and spectrally resolved multi-wavelength observations, that can provide insight into the star-dust geometry within these galaxies. 
Thus, many works rely on the use of scaling relations (\eg the [C$\textsc{ii}$]$\lambda158$ \um\ luminosity - SFR relation, \citealt{DeLooze2011}) or proxies \cite[\eg the infrared excess - UV continuum slope, referred to as the IRX-$\beta$ relation,][]{Smit2012} to derive the physical properties of galaxies. 
However, many of these relations are obtained from observations of the local Universe; one would expect that the physical conditions within high-redshift galaxies will differ substantially in the hot early Universe. 
Finally, observations of these high-z systems have shown that they are extremely clumpy \cite[\eg][]{Chen2023}, and hence generalising their properties from unresolved photometry will lead to incorrect inferences of their physical properties, such as stellar mass, SFR or dust attenuation \cite[\eg][]{Gimenez-Arteaga_2023}. 

A comprehensive way to explore the impact of star-dust geometry is through the use of hydrodynamical simulations of galaxy formation and evolution. 
These simulations typically associate dust with the distribution of metals in the gaseous medium. 
The last decade has seen the development of many calibrated physical models run over large periodic volumes to create representative galaxy samples [\eg\ \textsc{Eagle} \cite[]{schaye_eagle_2015,crain_eagle_2015}, \textsc{Illustris-tng} \cite[]{Marinacci2018,Naiman2018,Nelson2018,Springel2018,Pillepich2018}, \textsc{Simba} \cite[]{Dave2019}, \etc]. 
With sophisticated forward modelling techniques these simulations can now be compared to observed galaxies in the same observational space (\eg fluxes, luminosities or line equivalent widths). 
These advancements have led to many theoretical studies using simulations to understand the effect of dust on galaxy properties as well as the star-dust geometry \cite[\eg][]{Narayanan2018,Wilkins2018,Trayford2020,Shen2020,Lovell2022,Lower2022,Hsu2023}. 
They can help us explore and understand how galaxies -- which are complex objects with multiple spatially separated stellar populations of different ages (star formation histories) and metallicities (metal enrichment histories) -- contribute differently to the observed fluxes.

The recent launch of \jwst\ has ameliorated many of the drawbacks of space based telescopes for high-$z$ extragalactic 
%the void in astronomers' hearts :)
astronomy, due to its size, as well as its access to the observed frame near- and mid-infrared. 
The analysis of the data has already revealed interesting features about the high-redshift Universe \cite[\eg][]{Castellano2022,Ferreira2022,Naidu2022,Adams2023,Donnan2023,Harikane2023}, with some studies detecting massive galaxies with very high inferred values of dust obscuration as early as $z\sim10$ \cite[\eg][]{Rodighiero2023}. 
The spatial resolution of \jwst\ in the $1-5$ \um\ range also provides a new window for exploring resolved properties of high-z galaxies ($z\ge5$) in the rest-frame UV-optical for the first time without gravitational lensing \cite[\eg][]{Perez-Gonzalez2023}. 
This will allow us to probe the distribution of different stellar populations, as well as the intervening dust content. In order to fully comprehend these observations, it is imperative to construct suitable theoretical frameworks that facilitate their interpretation. 
In this context, it is timely to examine the variations in the observed UV-optical fluxes / luminosities resulting from the interplay between dust and the associated star-dust geometry. 
This work will complement other theoretical works in this space, with a focus mainly on the brightest and most massive galaxies in the high-redshift Universe. 

In this work, we explore the consequences that a realistic model for the star-dust geometry has on the observed properties of galaxies in the Epoch of Reionisation (EoR, $z\ge5$) using the First Light And Reionisation Epoch Simulations \cite[\flares,][]{Lovell2021,Vijayan2021}. 
\flares\ re-simulates a wide range of overdensities in the EoR, utilising the \eagle\ \cite[]{schaye_eagle_2015,crain_eagle_2015} simulation physics. 
By simulating a very wide distribution of overdensities in the EoR, \flares\ allows us to study the rare and bright objects, normally unavailable in periodic box simulations, due to their limited volume. 
This strategy also gives us access to a statistical sample of EoR galaxies \cite[see][]{Lovell2021}.  

This paper is structured as follows. In Section~\ref{sec:sim} we introduce the suite of simulations that we are using, our galaxy sample and explain briefly our spectral energy distribution (SED) modelling. 
We also introduce our methods to calculate the half-mass radii and observed SED for various viewing angles. In \S~\ref{sec: galvar} we explore the variation and consequence of the dust distribution within the simulated galaxies. 
In \S~\ref{sec: attcurve}, we also create toy galaxies to guide the interpretation of the impact of the variation of dust distribution in \flares\ galaxies, exploring its effect on the attenuation curve in \S~\ref{sec: attcurve} as well as on observed line ratios such as the Balmer decrement in \S~\ref{subsec:balmer} and the BPT diagram in \S~\ref{subsec:BPT}. 
In \S~\ref{subsec:los var} we look at the effect of various viewing angle on a few observed properties. We finally summarise our findings and present our conclusions in \S~\ref{sec:conc}. 

Throughout this work we assume a Planck year 1 cosmology \cite[$\Omega_{\mathrm{m}}=0.307$, $\Omega_{\Lambda}=0.693$, h $=0.6777$][]{planck_collaboration_2014} and solar metallicity, Z$_{\odot}=0.02$.

\section{Simulations and Modelling}\label{sec:sim}
\flares\ is a zoom-in hydrodynamical simulation suite of 40 regions, designed to study the evolution of galaxies hosted in different environments in the early Universe, ranging from extreme overdensities to extreme underdensities. These regions were chosen from a (3.2 comoving Gpc)$^3$ dark matter only (DMO) box, resimulated with full hydrodynamics using the AGNdT9 configuration of the \eagle\ \cite[]{schaye_eagle_2015,crain_eagle_2015} simulation physics to produce the galaxy population at $z\ge5$ with same resolution and cosmology as the reference \eagle\ simulation \cite[the physical model used to run the largest \eagle\ volume (100 cMpc)$^3$ , see Table 3 in][for the differences in the used parameters of the two models]{schaye_eagle_2015}. The AGNdT9 configuration produces similar mass functions to the Reference model but better reproduces the hot gas properties of groups and clusters \cite[]{barnes_cluster-eagle_2017}, with modifications that give less frequent, more energetic AGN feedback events. The model has been shown to be very successful in reproducing a number of observables in the low-redshift Universe not used in the calibration. The \flares\ philosophy is to test the \eagle\ physics model at high-redshift as well as forward model the simulated galaxies into the observed space. For this, we have deliberately selected extreme overdense regions (16), to obtain large sample of massive galaxies observable by telescopes in the high-redshift Universe. The properties of the high-redshift galaxies in \flares\ such as the UV-continuum slope evolution, galaxy stellar ages, metallicity have been shown to match the current observational constraints in a number of works \cite[\eg][]{Vijayan2021,Tacchella2022,Wilkins2022a}. The simulations have also been used to provide predictions for future observations in the high-redshift Universe \cite[\eg][]{Roper2022,D'Silva2023,Lovell2023,Wilkins2023b,Wilkins2023a}.
%\peter{This last sentence is unclear; it combines tests and predictions in a confusing way.}

To obtain representative distribution functions, we apply an averaging scheme that weights the different regions based on their overall abundance in the parent box, with the very overdense and underdense regions contributing the least. For more details we refer the interested readers to \cite{Lovell2021} where the region selection and weighting strategy are described in Sections 2.2 and 2.4, respectively. The main driver in using this suite of simulations is that it provides a statistically significant sample of galaxies in the EoR and thus enables us to better explore the evolution and properties of high-z galaxies.

\subsection{Galaxy Identification}\label{sec:sim.galid}
\begin{figure*}
    \centering
	\includegraphics[width=\textwidth]{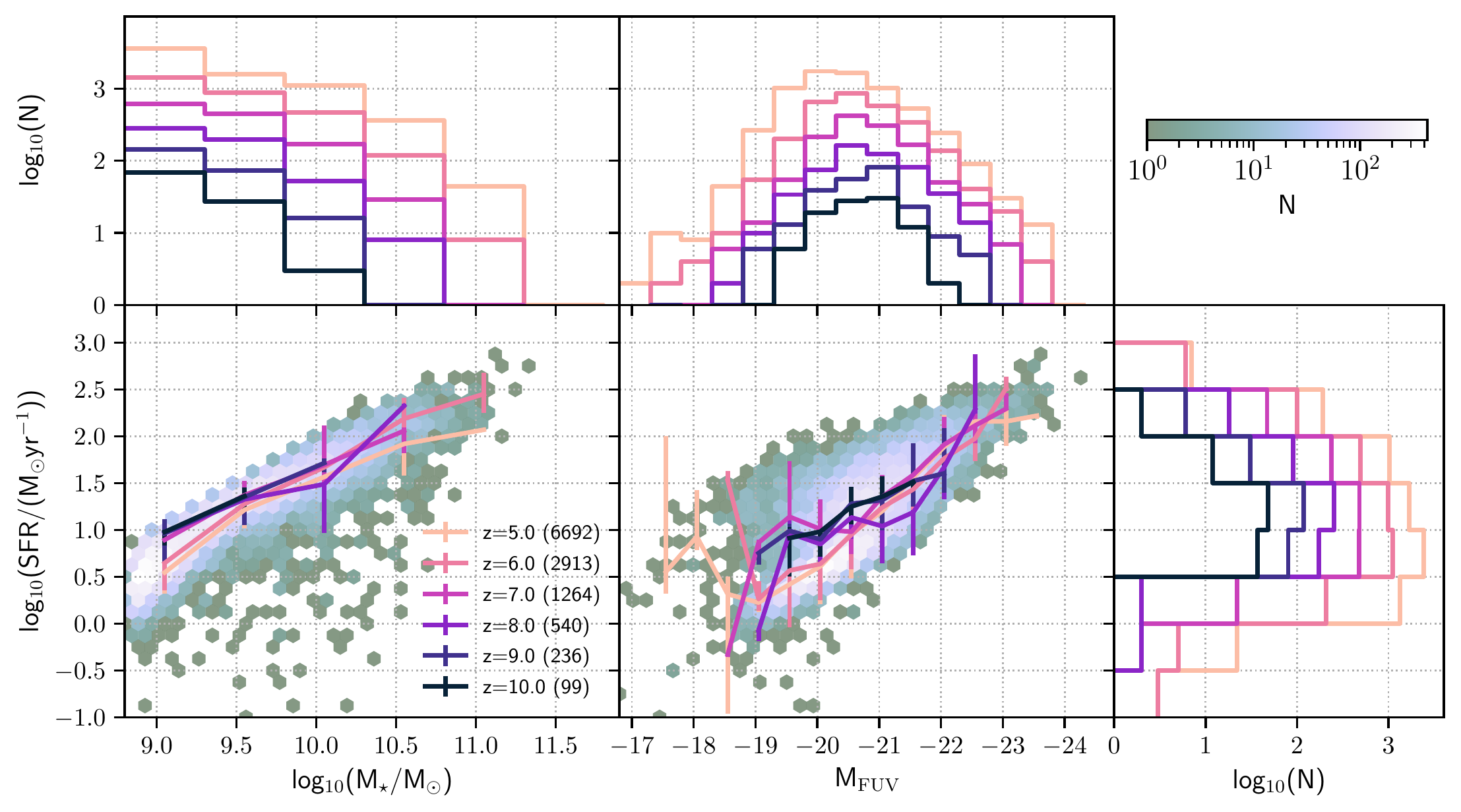}
	\caption{Shows the relationship of the galaxy stellar mass (left) and the galaxy UV luminosity (right) against their star formation rate (SFR) for $z \in [5,10]$. The coloured solid lines represent the weighted median of the relation at different redshifts, along with the $84^{th}-16^{th}$ percentile spread. Also shown is the histogram of the distribution of stellar mass and SFR in different bins for these redshifts. The total number of galaxies at these redshifts are indicated within brackets alongside the legend, with the colourbar indicating the number of galaxies within the hexbins. \label{fig: Mstar_Mfuv_sfr}}
\end{figure*}

Galaxies in \flares, similar to the fiducial \eagle\ model, are identified with the \textsc{subfind} \cite[]{springel_populating_2001,Dolag2009} algorithm, which runs on bound groups found by a Friends-Of-Friends \cite[\textsc{fof},][]{davis_evolution_1985} algorithm. The integrated galaxy properties like stellar mass, SFR or luminosities are calculated within a 30 physical kpc (pkpc) 3D aperture centered on the most bound particle of the self-bound structures, similar to previous \eagle\ and \flares\ works. For the redshifts of the \flares\ galaxies, this aperture ensures that we capture the majority of the mass/light belonging to each galaxy and remove any contribution from spurious distributions at larger radii. In this work, the quoted star formation rate (SFR) are measured using the mass of stars formed in the last 100 Myr. For the purpose of properly resolving and analysing the variation of different properties within galaxies, we assign a minimum threshold of 500 star particles for the galaxies in this study (corresponding to $\sim 10^{8.8}$M$_{\odot}$ in stellar mass). 

The left panel of Figure~\ref{fig: Mstar_Mfuv_sfr} shows the sample of galaxies used in this study in the redshift range of $5-10$. The figure plots the stellar mass as a function of the SFR of the galaxy. It can be seen that the sample extends over a stellar mass range of roughly $10^{8.8}-10^{11}$M$_{\odot}$, with the SFRs ranging from $0.1-1000$ M$_{\odot}$yr$^{-1}$.  

\subsection{Spectral Energy Distribution modelling}\label{sec:sim.sed}
We follow the same prescription adopted in \cite{Vijayan2021}, where we modelled the SED of each star particle in a galaxy, by treating them as a simple stellar population (SSP) based on their age and metallicity \cite[the smoothed metallicity of the star \ie\ SPH kernel weighted metallicity, see][]{EaglePdata2017}. The emission corresponding to a stellar population is then modelled using the the Binary Population and Spectral Synthesis (\textsc{bpass, v2.2.1}) SPS (stellar population synthesis) models \cite[]{BPASS2.2.1}. Throughout this work, we assume a Chabrier initial mass function \cite[IMF,][]{chabrier_galactic_2003}. We also associate each young stellar particle ($\lesssim 10$ Myr) with a surrounding H\textsc{ii} region powered by its LyC emission using the \textsc{cloudy (v17.02)} photoionisation code \cite[]{Cloudy17.02}; same as the photo-ionisation model described in \cite{Wilkins2020}.

Since our simulations do not model the formation and destruction of dust, the dust attenuation is modelled by converting the line-of-sight metal column density along the z-axis of the simulation (the adopted viewing angle, thus giving a random orientation for each galaxy) to a dust column density. This conversion is done by adopting a dust-to-metal ratio \cite[using the fitting function for the dust-to-metal (DTM) ratio from][equation 15]{Vijayan2019} for the galaxy, implemented as follows: 
\begin{gather}
	\tau_{\textrm{ISM,V}}(x,y) = \mathrm{DTM}\,\kappa_{\textrm{ISM}}\,\Sigma\,(x,y),
\end{gather}	
where $\tau_{\textrm{ISM,V}}(x,y)$ and $\Sigma\,(x,y)$ are the V-band (550nm) optical-depth and integrated metal column density respectively of the intervening diffuse ISM along the line-of-sight at position (x,y) of the star particle. $\kappa_{\textrm{ISM}}=0.0795$, is a proportionality constant, chosen to match the rest-frame far-UV (1500\AA) luminosity function to the observed UV luminosity
function from \cite{Bouwens2015} at $z=5$.

We also assume a contribution to the dust attenuation from the stellar birth clouds, which dissipate over a timescale of 10 Myr \cite[]{CF00}. For the stellar birth clouds, we scale their dust content with the smoothed star particle metallicity, described as follows,
%\peter{How does that give a constant dust-to-metal ratio, unless you smooth over the whole galaxy, which would be weird?}. 
\begin{gather}
\tau_{\textrm{BC,V}}(x,y) = 
	\begin{cases}
		\kappa_{\textrm{BC}} (\textrm{Z}_{\star}/0.01)\, & \text{t} \leq 10^7\textrm{yr}\\
		0\, & \text{t} > 10^7\textrm{yr}\:,
	\end{cases}
\end{gather}
where $\tau_{\textrm{BC,V}}(x,y)$ is the V-band optical-depth due to the stellar birth-cloud and \textrm{Z}$_{\star}$ is the smoothed metallicity of the young stellar particle. $\kappa_{\textrm{BC}}=1.0$, similar to $\kappa_{\textrm{ISM}}$, is a normalisation factor, chosen to to match the UV-continuum slope observations from \cite{Bouwens2012,Bouwens2014} at $z=5$ and the [O\textsc{iii}]$\lambda4959,5007$ + H$\beta$ equivalent width distribution at $z=8$ from \cite{deBarros19_OIIIHbeta}. The value also encapsulates the dust-to-metal ratio in the stellar birth clouds. By fixing the values of $\kappa_{\textrm{ISM}}$ and $\kappa_{\textrm{BC}}$ for all redshifts, we assume there is no evolution in the general properties of the dust grains in galaxies such as the average grain size, shape, and composition.

We link the optical-depth in the V-band to other wavelengths using a simple powerlaw relation:
\begin{equation}\label{eq:tau_lambda}
	\tau_{\lambda} = (\tau_{\textrm{ISM}} + \tau_{\textrm{BC}}) \times\,(\lambda/550\textrm{nm})^{-1}\:.
\end{equation}
This produces an extinction curve for a stellar particle flatter in the UV than the Small Magellanic Cloud curve \citep{Pei1992}, but not as flat as the \cite{Calzetti2000} curve. We refer the interested readers to Section 2.3, 2.4 and Appendix A of \cite{Vijayan2021} for an in-depth discussion of the photometry generation in \flares.  Our modelling differs from using a uniform screen of dust across a galaxy (usually assumed in SED fitting of observed galaxies), with spatially distinct stellar populations experiencing differing amounts of dust attenuation, thus a more realistic model for dust.

The right panel of Figure~\ref{fig: Mstar_Mfuv_sfr} shows the UV luminosity of the galaxies obtained using the described forward modelling, plotted against the galaxy SFR for $z\in[5,10]$. From the histogram of the distribution of the UV luminosity, it can be seen that the sample contains more number of brighter ($<-22$ Mag) and fainter ($>-19$ Mag) galaxies as one goes towards lower redshifts, similar to the SFR distribution. The lower SFR cut at $z=10$ is due to our star particle selection limit. The galaxies with similar SFR have different UV luminosities due to the variation in the amount of dust. Moving down in redshift, the distribution extends towards lower luminosities ($>-19$ Mag) due to the build-up of dust as well as galaxies with older stellar populations, while the presence of very massive young star forming galaxies extends the distribution at the bright end ($<-22$ Mag).

\subsection{Half-mass radii}\label{sec:sim.hmr}
Hydrodynamical simulations like \flares\ provide information on the spatial distribution of stars and gas within a galaxy. We can utilise this to calculate the mass distribution within the galaxy and thus derive the half-mass radii for either of these components. To calculate the half-mass radii we compute the distance of each star/gas particle within a galaxy from its centre of potential. By ordering them in increasing distance one can find the radius that encloses half the mass in that component. The half-mass radii will be used to normalise the radial variation in observed galaxy properties in \S~\ref{sec: galvar}.

Galaxy sizes in \flares\ have been extensively discussed in \cite{Roper2022,Roper2023}; who find that galaxies at the massive end in the early Universe have most of their star formation localised to extremely compact central cores. A main reason for these compact cores are due to the gas densities required for star formation in the early Universe being very high, because of the very low gas phase metallicities. The \eagle\ model also defines a gas critical density at which a star particle can be formed, which depends inversely on the gas phase metallicity, thus driving the critical density higher in the early Universe. And once star formation is triggered in these regions, they become quickly enriched by metals, starting a process of runaway star formation. Thus these compact cores are very metal rich and suffer from extreme dust attenuation. This causes these galaxies to have observed (\ie dust attenuated) UV sizes that are significantly larger than the intrinsic (\ie no obscuration by dust) UV sizes. These compact systems evolve to larger intrinsic as well as observed UV sizes due to the surrounding gas becoming enriched at later times. For more details about the size evolution of the \flares\ galaxies, interested readers are directed to the referred works.

\subsection{Re-projection}\label{sec:sim.reproject}
We use the Hierarchical Equal Area isoLatitude Pixelisation (\texttt{HEALPix}\footnote{\url{http://healpix.sourceforge.net/}}) scheme as implemented in \texttt{healpy}\footnote{\url{https://github.com/healpy/healpy/}} to pixelize the line-of-sight angles. We assume $N_{\mathrm{side}}=8$ yielding 768 individual lines-of-sight. We re-project the star and gas particles of the galaxy along these lines-of-sight, with the viewing angle along the z-axis as before. We then calculate the dust-attenuated SED.
This allows us to explore the observed properties of galaxies as a function of viewing angle (\eg\ see Figure~\ref{fig: healpy_plots}). In \S~\ref{subsec:los var} we use these re-projected values to quantify the effect of different viewing angles on the observed galaxy properties.  
\section{Variation of UV attenuation within galaxies}\label{sec: galvar}
\begin{figure*}
	\centering
	\includegraphics[width=\textwidth]{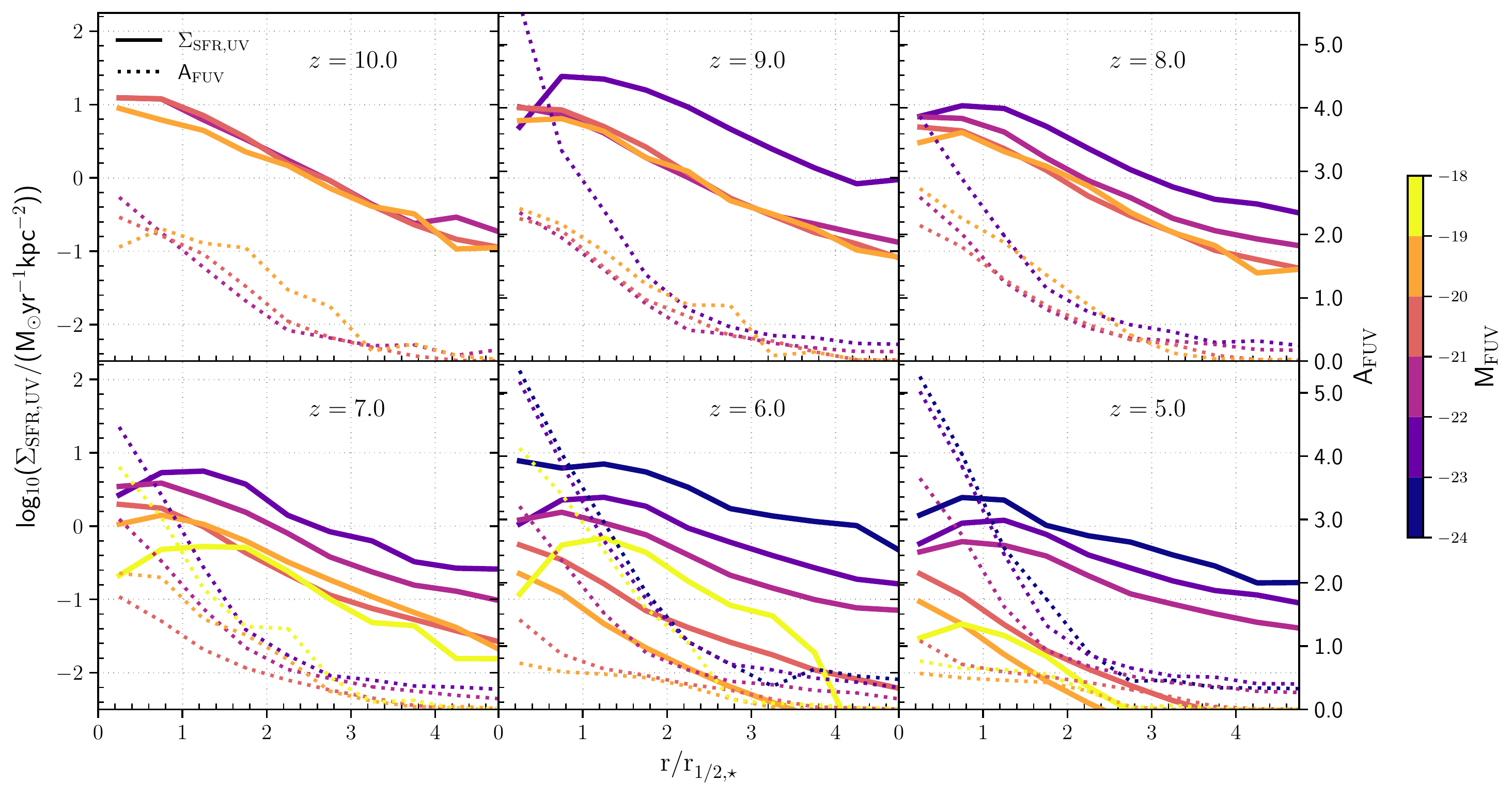}
	\caption{Variation in the UV star formation surface rate density (solid lines) and the UV-attenuation (dotted lines) with distance from the galaxy centre, in units of the half-mass radius. Only M$_{\mathrm{FUV}}$ bins with $\ge10$ galaxies are plotted. There is only negligible change if we use an SFR averaging timescale of 10 Myr instead of 100 Myr. \label{fig: sfratt_rad_mstar}}
\end{figure*}
In this section we explore how dust attenuation varies within individual galaxies. While galaxies can be characterised by a global attenuation, it is clear that they are complex objects with different regions experiencing varying degrees of dust attenuation. The line-of-sight dust attenuation model implemented in \flares\ allows us to explore this variation. We begin in \S~\ref{sec:galvar:uvlf} by exploring how dust attenuation and the star formation traced by the UV emission (\ie the unobscured SFR, denoted by SFR$_{\mathrm{UV}}$) correlates with distance from the galaxy centre and its implications. From here onwards, to make the distinction from SFR$_{\mathrm{UV}}$, we will denote the total SFR of the galaxy as SFR$_{\mathrm{Total}}$, which includes contribution from both obscured and unobscured star formation. We then explore in \S~\ref{sec:galvar:var_spop} the spread in the degree of attenuation within individual galaxies of different star particles.

\subsection{Radial variation and the luminosity function}\label{sec:galvar:uvlf}
\begin{figure*}
	\centering
	\includegraphics[width=0.95\textwidth]{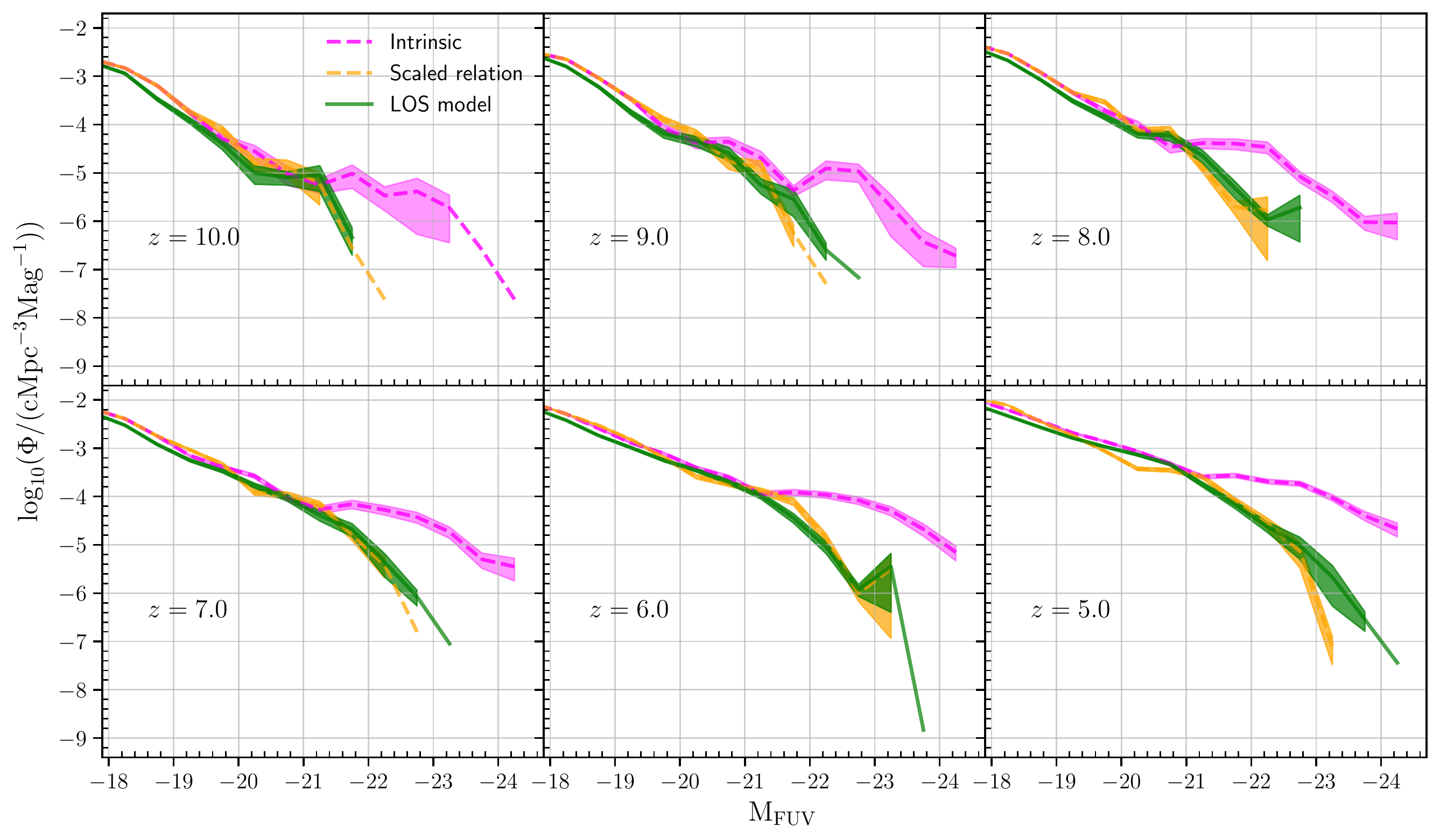}
	\caption{The intrinsic (dotted blue) UV LF and the observed UV LF for the line-of-sight model (solid green) and the one obtained using equation~\protect\ref{eq: Auv_linear} (dashed orange) for the \flares\ galaxies in $z\in[5,10]$ are shown. The shaded region denotes the $1-\sigma$ Poisson errors on the numbers, and is only shown for bins with $\ge5$ galaxies. Note that the star particle cut mentioned in \S~\ref{sec:sim.galid} is not enforced here to preserve the full shape of the UV LF. \label{fig: uvlf}}
\end{figure*}
Using the half-mass radii calculated in \S\ref{sec:sim.hmr} we describe how the average UV attenuation within galaxies varies with distance from the centre. 
In Figure~\ref{fig: sfratt_rad_mstar}, we plot the median variation with radius of the UV star formation rate surface density ($\Sigma_{\mathrm{SFR,UV}}$), and the attenuation in the far-UV (in magnitudes, A$_{\mathrm{FUV}}$), for galaxies at $z\in[5,10]$. 
We compute SFR$_{\mathrm{UV}}$ in a galaxy by defining the observed-to-intrinsic fraction of UV emission, $f_{\mathrm{esc}}$, using
\begin{gather}\label{eq: f_esc}
	f_{\mathrm{esc}} = \frac{L_{\mathrm{FUV}}^{\mathrm{Observed}}}{L_{\mathrm{FUV}}^{\mathrm{Intrinsic}}} = 10^{-A_{\mathrm{FUV}}/2.5}.
\end{gather}
We can then define
\begin{gather}\label{eq: SFRuv}
    \mathrm{SFR}_{\mathrm{UV}} = f_{\mathrm{esc}}\times\mathrm{SFR}_{\mathrm{Total}},\\
    \Sigma_{\mathrm{SFR,UV}} = \mathrm{SFR}_{\mathrm{UV}}/(\pi (r_{o}^2 - r_{i}^2)), 
\end{gather}
where $r_{o}$ and $r_{i}$ are the radii of the outer and inner bins (in kpc) respectively, considered for calculating the surface area. We reiterate here that $f_{\mathrm{esc}}$, in the context of this work, does not refer to the escape fraction of ionising photons from H$\textsc{ii}$ regions, but it refers to the fraction of UV emission that is observed or unobscured. In Figure~\ref{fig: sfratt_rad_mstar}, we have normalised the radial bins with respect to the stellar half mass radii (r$_{1/2,\star}$).

From Figure~\ref{fig: sfratt_rad_mstar}, it can be seen that the attenuation of the stellar light is highest in the centre of the galaxies, rapidly decaying towards the outskirts.
This is not surprising considering that stars are formed first in the dense central cores, with the subsequent generation of stars forming from gas enriched by this population. 
This inside-out growth of star formation within the galaxy builds up higher dust surface densities within the inner parts of the galaxy, leading to higher dust attenuation in the centres. 
At $z=10$, the median radial variation of $\Sigma_{\mathrm{SFR,UV}}$ for different galaxies with different observed FUV luminosities are almost the same. 
This is due to the galaxies in the most luminous bin ($\le-20$ Magnitude) having higher UV obscuration in the centre. The galaxies on the fainter side at $z=10$ show almost uniform dust obscuration within the inner $2\times$r$_{1/2,\star}$. This is also accompanied by slightly lower $\Sigma_{\mathrm{SFR,UV}}$ in the inner regions. 
Moving to lower redshifts, the brightest luminosity bins ($\le-22$ Magnitude) show significant amount of dust obscuration within the central region\footnote{except at $z\in[6,7]$ where the highest bin ($-19$ -- $-18$ in M$_{\mathrm{FUV}}$) contains extremely dust obscured massive galaxies}. 
At $z=5$ the brightest galaxies ($> -21$ Magnitude) are still intensely star-forming with most of the activity being highly-obscured in the centre. This is due to highly dust enriched central regions in these galaxies \cite[see][]{Roper2022}.

However, the radial decline in the amount of UV star formation is not as dramatic as the UV attenuation, with the figure clearly indicating there is a significant amount of UV luminosity that is coming from the outskirts, especially for the brightest galaxies. This effect is also reflected in the UV size evolution of galaxies across redshift, with the intrinsic sizes as much as 50 times smaller than the observed sizes \cite[see][]{Roper2022,Roper2023}. 
In short, galaxies at the bright end of the UVLF have started forming stars away from the central dust rich core towards the outskirts. However, since the gas in the outskirts is less enriched, there is thus less dust attenuation. This phenomenon makes these galaxies brighter than expected compared to the galaxies in the lower luminosity bins.
Thus, in \flares, for the bright galaxies we are seeing a non-linear relationship (in the magnitude space) between the UV luminosity and the UV attenuation.

To probe the effect of this extended star formation and the resultant lower attenuation in these galaxies have  
% the described phenomenon 
on the shape of the UV LF at the bright end, we perform the following simple exercise. We convert the intrinsic UV LF of the \flares\ galaxies to the dust attenuated value using a linear scaling relation between the two, which is commonly used in literature \cite[\eg][]{Smit2012,Tacchella2013}. The scaling relation assumes that there is a one-to-one relation between the intrinsic UV luminosity and dust attenuation, which means that it would fail to properly capture the effect of unobscured star formation in the outskirts. This is done by following the parameterisations for the mean UV-continuum slope and the UV attenuation:
\begin{equation}
	\langle\beta(z,M_{\mathrm{UV}})\rangle = \frac{\mathrm{d}\beta}{\mathrm{d}M_{\mathrm{UV,int}}}(z)[M_{\mathrm{UV}} - M_{0}] + \beta_{M_{0}}(z),
\end{equation}
which feeds into
\begin{equation}
	\langle A_{\mathrm{UV}}(z) \rangle = 4.43 + 0.79\mathrm{ln}(10)\sigma_{\beta}^2 + 1.99\langle\beta(z,M_{\mathrm{UV}})\rangle,
\end{equation}
and 
\begin{equation}\label{eq: Auv_linear}
    M_{\mathrm{UV,att}} = M_{\mathrm{UV,int}} + \langle A_{\mathrm{UV}}(z) \rangle.
\end{equation}
where $M_{\mathrm{UV,int}}$ is the intrinsic UV luminosity, $M_{\mathrm{UV,att}}$ the dust attenuated UV luminosity. $M_{0}=-19.5$, $\sigma_{\beta}=0.34$ and the values of $\frac{\mathrm{d}\beta}{\mathrm{d}M_{\mathrm{UV}}}(z)$ and $\beta_{M_{0}}(z)$ are taken from Table 1 of \cite{Tacchella2013}\footnote{in case of $z>8$, we use the $z=8$ values and negative $\langle A_{\mathrm{UV}}(z) \rangle$ is set to zero}. These are in turn from \cite{Bouwens2012,Bradley2012}. 

Figure~\ref{fig: uvlf} shows the intrinsic (dotted blue line) UV LF, with the dashed orange line showing the attenuated UV LF using Equation~\ref{eq: Auv_linear} (linear model, in figure). We also plot the observed UV luminosity using the line-of-sight dust attenuation model (the default model for the \flares\ galaxies described in \S\ref{sec:sim.sed}; LOS model in figure) as the solid green line.  
As can be seen from the plot, the two models produce similar number densities for the fainter galaxies, while from $z\le9$, the galaxy number densities are extended to higher luminosities compared to the scaled model. 
This extension to higher luminosities is not strongly visible as we go to higher redshifts due to \flares\ missing the extreme bright galaxies. However, it is clear that across redshifts, the brightest galaxies have less global dust attenuation in the LOS model, and thus the number densities have been pushed to higher values \footnote{also see Figure 9 in \cite{Vijayan2021} for the \flares\ UV attenuation relation} 
than the scaled model. This is due to the latter's simpler parameterisation of the relationship between intrinsic UV luminosity and dust attenuation, \ie a linear relation between attenuation, $\beta$ slope and UV luminosity. For the \flares\ galaxies, the relationship between these values deviate from a linear relationship and becomes flat at the bright end. This phenomenon is partly responsible for the \flares\ UV LF exhibiting the double power-law (DPL) shape measured in \citep{Vijayan2021}:
the double Schechter shape of the intrinsic UV luminosity function in \flares, when attenuated (or observed), becomes a DPL.

An implication of the above is that the use of a linear scaling between the UV attenuation and luminosity will over-predict (since there is agreement at the faint-end) the UV attenuation at the bright end of the UV luminosity function. Thus the estimated intrinsic UV luminosity will be higher. Also the predicted stellar mass and star formation rates will be wrong due to widely varying mass-to-light ratios across the galaxy.
Thus it is important to take into account such biases when interpreting derived intrinsic values from observed UV luminosity, when the derived values are estimated with the use of UV-to-SFR conversion factors or constant mass-to-light ratios. 

\subsection{Variation across stellar populations}\label{sec:galvar:var_spop}
\begin{figure*}
	\centering
	\includegraphics[width=\textwidth]{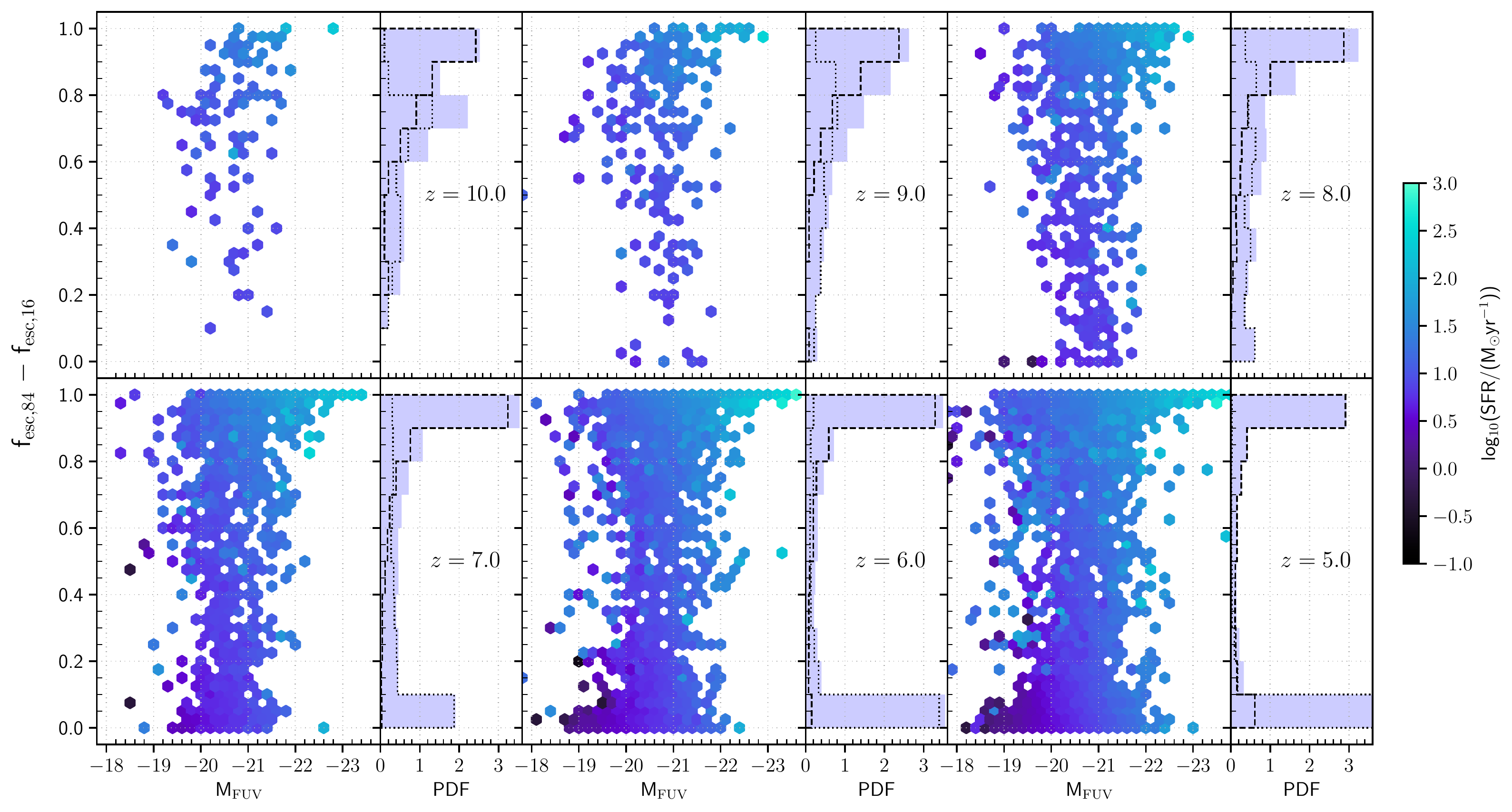}
	\caption{Shows the variation of the dust obscuration in the UV within galaxies as a function of the observed UV luminosity. The spread is described as the difference between the 84th and 16th percentile of the UV attenuation of stars within the galaxy. The hexbins have been coloured by the median SFR$_{\mathrm{Total}}$ of the galaxies in the bin. The plotted distribution function alongside shows the normalised counts of galaxies in different bins. The dashed line is the distribution of fraction of galaxies with their star formation rate higher than the median value at that redshift, with the dotted line showing the fraction of galaxies below the median. \label{fig: fesc_distr_z5_10}}
\end{figure*}
In this subsection we will explore how much scatter there is in the dust attenuation seen by different star-forming regions (or particles in the simulation). Figure~\ref{fig: fesc_distr_z5_10} shows the spread in the UV attenuation (or its proxy, $f_{\mathrm{esc}}$ in equation~\ref{eq: f_esc}, now for individual star particles; where $f_{\mathrm{esc}}\xrightarrow{}0$ indicates a fully dust obscured star forming region, and vice versa)
% \peter{escape fraction, surely?} 
in the UV for $z\in[5,10]$. Here we describe the scatter within a galaxy, by the difference between the 84th and 16th percentile of the different $f_{\mathrm{esc}}$ values of star particles, \ie\ $f_{\mathrm{esc,84}}-f_{\mathrm{esc,16}}$. 
A high value of $f_{\mathrm{esc,84}}-f_{\mathrm{esc,16}}$ ($\xrightarrow{}1$) is indicative of extremely inhomogeneous dust distribution within the galaxy. 

Figure~\ref{fig: fesc_distr_z5_10} shows this scatter for different galaxies in the redshift range $[5,10]$ as a function of the observed UV luminosity, with the hexbins coloured by the median SFR$_{\mathrm{Total}}$ within that bin. 
We also plot alongside the probability distribution function (PDF, normalised by the bin width) of $f_{\mathrm{esc,84}}-f_{\mathrm{esc,16}}$ for our galaxy sample. 
The figure shows that there is a huge spread within most galaxies at $z=10$ (see PDF), with the number of galaxies with small spread in the the $f_{\mathrm{esc}}$ values increasing with decreasing redshift. 
At all redshifts, in \flares\ there is a large spread in $f_{\mathrm{esc,84}}-f_{\mathrm{esc,16}}$ that is correlated neither with the galaxy UV luminosity (as can be seen directly from Figure~\ref{fig: fesc_distr_z5_10}) nor the stellar mass. 
We have also looked at the contribution to this scatter due to young stars still embedded in their birth clouds, and found it to be negligible. 
From the hexbin colours, it can be seen that there is a weak correlation with the SFR$_{\mathrm{Total}}$ of the galaxy at $z=10$, with the correlation increasing with decreasing redshift. I
f we concentrate on the highly star forming galaxies in \flares, by splitting them using the median SFR$_{\mathrm{Total}}$, it can be seen that they predominantly occupy the upper part of $f_{\mathrm{esc,84}}-f_{\mathrm{esc,16}}$, while the the lower half consists of less star forming galaxies. 
A similar pattern is also seen if one were to replace SFR$_{\mathrm{Total}}$ with UV attenuation, however for the rest of this work, we use SFR$_{\mathrm{Total}}$ to quantify this spread.

This observed spread is mostly a result of the complex star-dust geometry within the galaxy, thus having mix of stars with highly obscured and unobscured sightlines. 
The observed spread is seen to be correlated with the galaxy SFR in \flares. 
The correlation implies that highly star forming galaxies have more regions that have sightlines suffering different levels of dust attenuation. 
This is not unexpected, since we also saw in \S\ref{sec:galvar:uvlf} that extreme UV-bright galaxies have star formation that is less obscured, due to stars forming outside the dust-rich core. 
This result is robust in the mass range we are investigating, due to the conservative particle number cut employed in this study, which implies that it is highly unlikely that the dust distribution is poorly sampled (creating holes within the distribution).

Another interesting feature we observe is that with decreasing redshift there are more galaxies that start to populate the region with $f_{\mathrm{esc,84}}-f_{\mathrm{esc,16}} \sim 0$. 
This is indicative of galaxies with a more homogeneous dust distribution, or with very low levels of dust attenuation (the spread can still be significant if the attenuation is very low). In these cases, the application of a homogeneous screen model will be suitable. There are two reasons for the surge in the number of such galaxies. One of the reason for more galaxies is that in general as one goes to lower redshift, we have more galaxies in the simulation fulfilling our selection criteria. Secondly, galaxies at higher redshift are expected to have undergone a more bursty phase of star formation \cite[]{Wilkins2023b}, and thus more likely to exhibit significant variation in the dust attenuation across the galaxy (due to the seen corelation with SFR). However, as these galaxies move to lower redshifts, their star formation rates are expected to gradually decline and undergo fewer frequent bursts. Hence they are expected to have more homogeneous distribution of dust along the line-of-sight (also see \S~\ref{sec: attcurve}).

Recently, \cite{Lower2022} implemented an additional parameter in SED fitting, $f_{\mathrm{unobscured}}$, to take into account contribution from stars which lie across sightlines which suffer from negligible dust-attenuation. The model was implemented in the SED fitting code \texttt{Prospector} \cite[]{Prospector2021} and was used to fit the observed fluxes (19 bands, including \textit{GALEX}, \textit{HST}, \textit{Spitzer}, and \textit{Herschel}) of mock galaxies at $z=0$ from the $\textsc{Simba}$ simulations generated using the radiative transfer code \texttt{Powderday} \cite[]{Narayanan2021}. The results from the study showed that incorporating an extra parameter to allow for unobscured sightlines provide better constraints on derived values such as stellar mass and star formation rate. Their results are in agreement with ours, showing the value of incorporating an unobscured parameter in SED fitting.

\section{Attenuation Curve}\label{sec: attcurve}
\begin{figure*}
	\centering
	\includegraphics[width=0.85\textwidth]{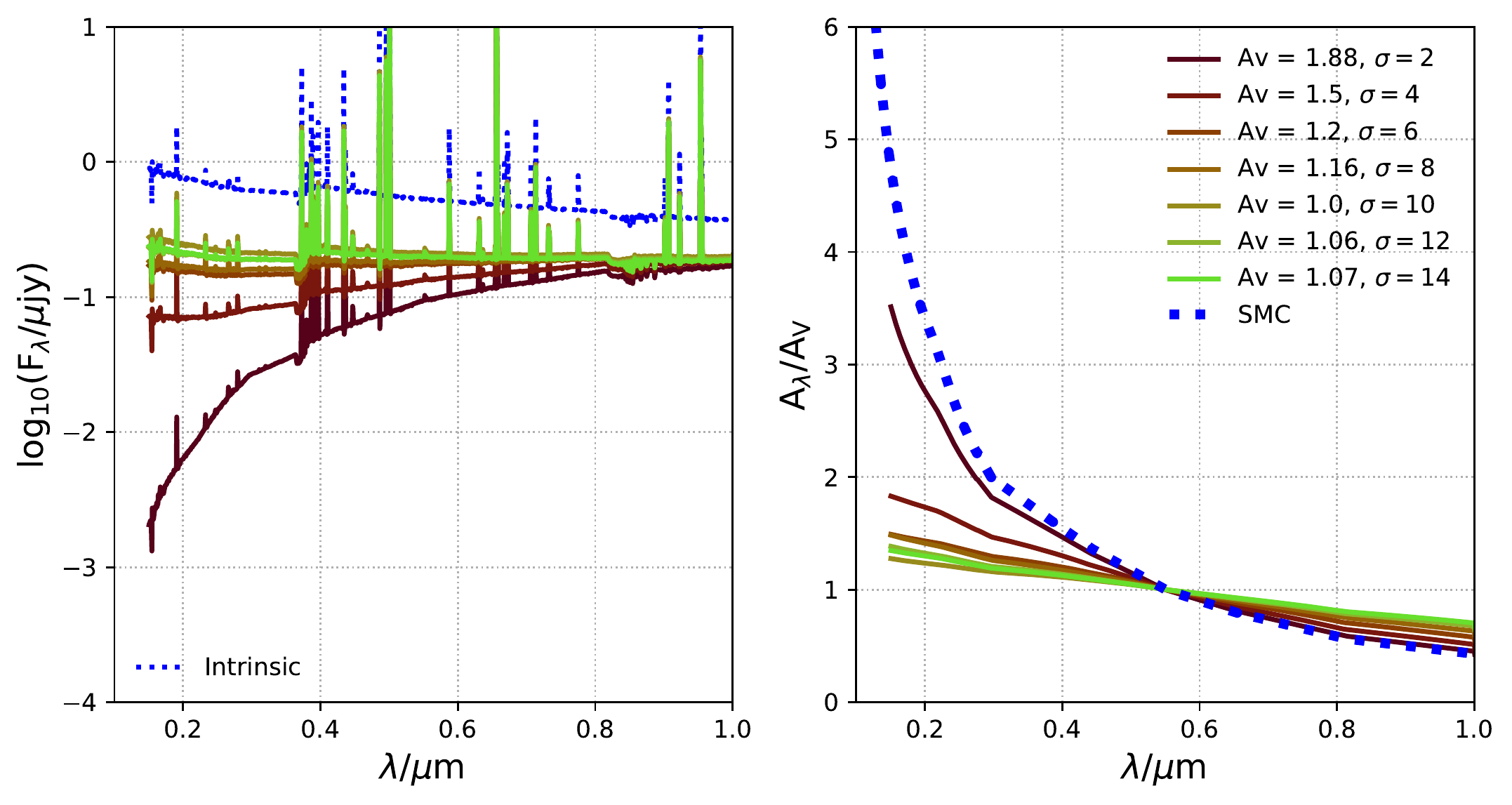}
	\caption{\textbf{Left}: The intrinsic (blue, dotted) and observed (for different spread in attenuation, in coloured solid) spectrum of the galaxy by summing up the contribution from the different star-forming clumps made with SED fitting code \texttt{Bagpipes}. The regions were all assigned the same stellar ages with a single burst SFH, metallicity, ionisation parameter and mass (see \S\ref{sec: attcurve}). Each clump follows an SMC dust attenuation law, where we vary the line-of-sight attenuation by varying the A$_{\mathrm{V}}$ magnitude. \textbf{Right}: Attenuation curves of the resultant galaxy, which is shown in different colours, obtained varying the dust attenuation of the individual clumps. The green dotted line is the input SMC attenuation curve. \label{fig: pipegal}}
\end{figure*}
\begin{figure*}
        \centering
	\includegraphics[width=\textwidth]{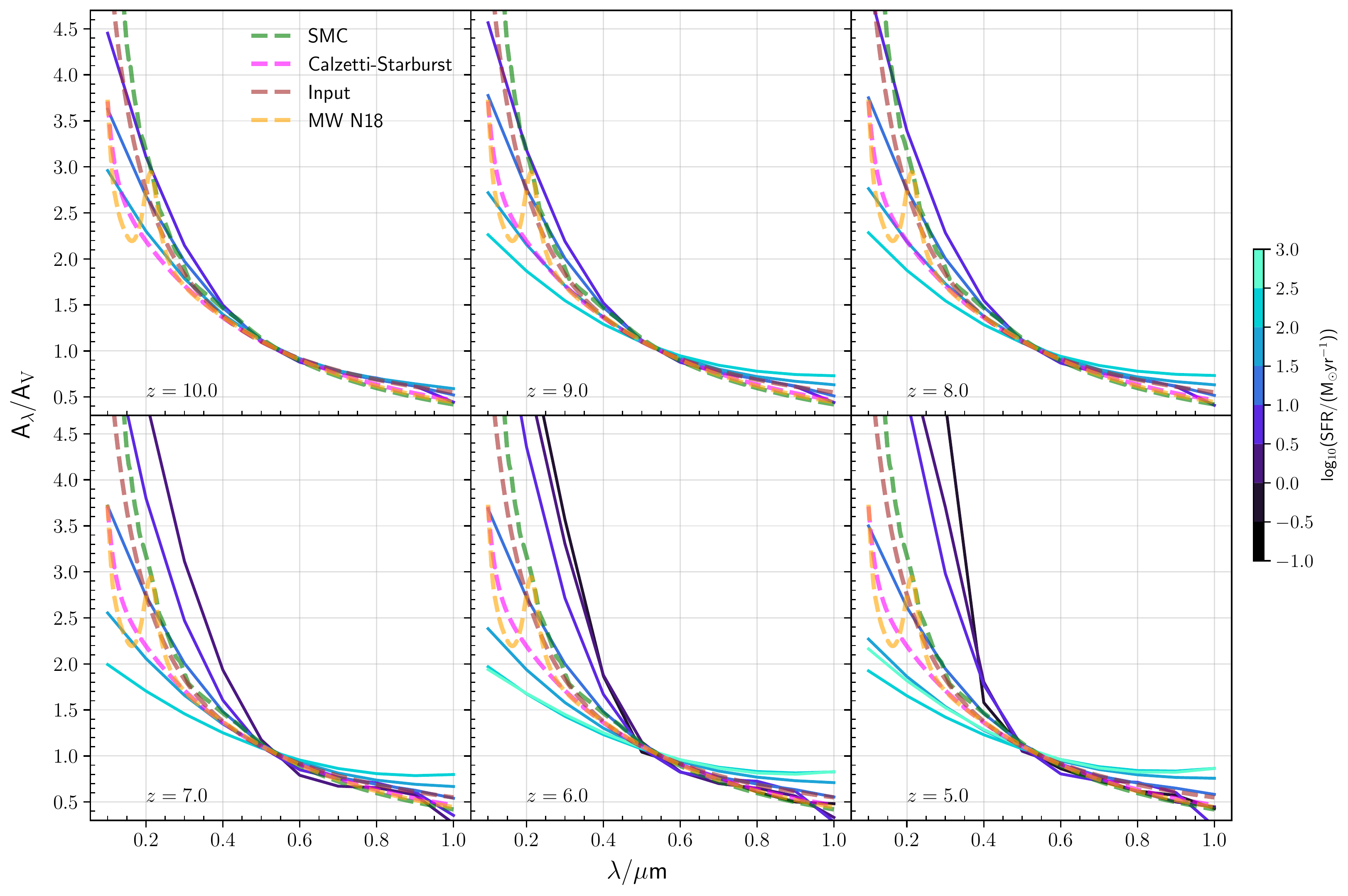}
	\caption{Galaxy attenuation curve for the \flares\ galaxies in different bins of SFR$_{\mathrm{Total}}$. The different panels shows the attenuation curve for $z\in[5,10]$. The curves have been smoothed and the emission lines removed for better visual clarity. We also plot alongside attenuation curves from literature: SMC \protect\citep[]{Pei1992}, Calzetti-Starburst \protect\citep[]{Calzetti2000} and the Milky Way curve from \protect\citet[][MW N18]{Narayanan2018} as well as the \flares\ input attenuation curve.} \label{fig: att_curve}
\end{figure*}
\begin{figure}
        \centering
	\includegraphics[width=0.87\columnwidth]{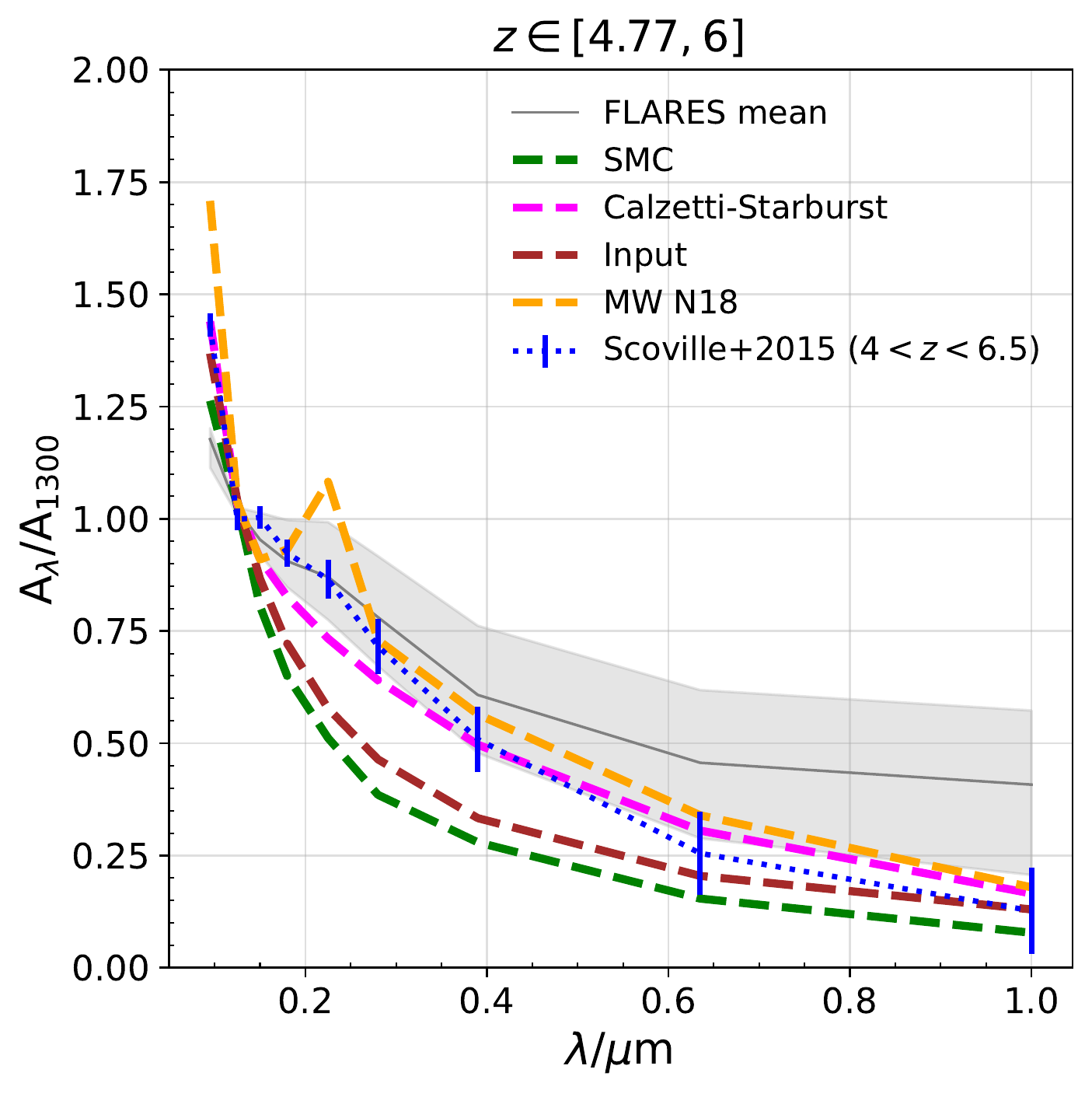}
	\caption{Galaxy attenuation curve with similar selection as the \protect\cite{Scoville2015} galaxies of I$_{\mathrm{AB}}<25$. We combine our galaxies in the redshift range $[4.77,6]$, with shaded region showing the $16^{th}-84^{th}$ percentile spread. All the other attenuation curves plotted from literature are from the low-redshift Universe.\label{fig: att_curve_IAB}}
\end{figure}
The observed spectra of galaxies are a result of different contributions from the stars of different ages and metallicities that are inhomogeneously distributed in a clumpy soup of dust. A number of steps are required to model the intrinsic spectrum of galaxies \cite[see][for a detailed review]{Walcher2011_review,Conroy2013_review}. Dust attenuation curves describe how the intrinsic emission (from stellar and nebular emission) from a galaxy is modified by dust at different wavelengths. 
An understanding of these attenuation curves is essential for accurately determining physical properties of galaxies, through SED fitting of the observed galaxy photometry or spectrum. Therefore, they play a pivotal role in galaxy formation and evolution studies.

We already saw in the previous section that the galaxies in \flares\ have a wide distribution of dust attenuation across different star particles. In order to understand the effect this variation has on the attenuation curve of galaxies as well as on the conclusions drawn from the observed SEDs, we create a controlled experiment. This will help us better understand the observed behaviour exhibited by the \flares\ galaxies in this and the next section.
% In order to demonstrate this, 
For this we build toy galaxies at $z=5$ in the SED fitting code \texttt{Bagpipes} \citep{Carnall2018} with the 2016 updated version of BC03 SPS \cite[]{BC03} library, modelling it as comprising of 50 star forming clumps. The resulting galaxy has a composite SED obtained by summing up the SEDs of the individual clumps. To generate the SEDs for these star forming regions we assume 
% \peter{Be consistent in tense -- you have siwtched from past to present} 
a single burst star formation history (SFH) with the same stellar age (5 Myr), metallicity (0.1 Z$_{\odot}$), mass of stars formed (10$^7$ M$_{\odot}$) and logU (ionisation parameter, -3). We assume a SMC type dust law, and change the line-of-sight dust attenuation by varying the attenuation in the V-band (A$_{\mathrm{V}}$ magnitude) for the clumps. Thus the intrinsic spectrum of the galaxies are the same, however, the observed spectrum is different. 
It should be noted that in general SED fitting codes assume the same extinction or attenuation (commonly referred to as a dust screen, with the choice of curve determining how porous the screen is) for all the different stellar components (with multiple ages and metallicities), with extra attenuation only for young stars which are still embedded within their birth clouds. We also produce these clumps with no added contribution from nebular emission or extra dust from young stars and find similar conclusions to the results presented later.

We assess the impact of the variation of dust attenuation across the galaxy on the resultant attenuation curve by randomly drawing the A$_{\mathrm{V}}$ from a normal distribution with a mean ($\mu$) of $2$ and standard deviation ($\sigma_0$) of $0.3$. We quantify the spread within the galaxy by varying the standard deviation from $2\sigma_0-14\sigma_0$ \ie
\begin{equation}
    A_{\mathrm{V},i} = \mu + \mathcal{N}(0,n\sigma_{0}),
\end{equation}
where A$_{\mathrm{V},i}$ is the A$_{\mathrm{V}}$ value of the i$^{th}$ clump and $n\in[2,14]$, quantifying the spread. Any negative A$_{\mathrm{V},i}$ value is changed to $0.01$. The reason for choosing this particular mean A$_{\mathrm{V}}$ along with wide values of the spread in attenuation, is to replicate the variations in the dust attenuation that are similar to the galaxies in the \flares\ simulation. This will become increasingly clear from the following discussions.  

In Figure~\ref{fig: pipegal} we show the result of our experiment. The left subplot shows the intrinsic SED of the galaxy in dotted blue line, with the various coloured solid lines showing the observed SED after assigning the different values of dust attenuation for the star forming clumps forming the galaxy. The right subplot shows resultant attenuation curves for the various $n\,\sigma_{0}$ choices in solid coloured lines. The input SMC curve is also shown as the green dotted line. Emission line features are not seen in the resultant curves since all the star forming clumps have the same age and metallicity. It can be clearly seen that with increasing value of $n$ (\ie more spread in dust attenuation), the resultant attenuation curve starts to deviate from the input SMC curve (one recovers the input SMC curve, if the different clumps in the toy galaxy were provided with similar values of A$_{\mathrm{V}}$ magnitude). Higher values of $n$ make the curve significantly flatter than the input curve, and at the longer wavelength end the normalisation is higher as well. Previous studies have also shown that in a clumpy medium, one can go from SMC type dust law or grain composition to produce a Calzetti like law which is significantly greyer than the SMC dust law \cite[\eg][]{Witt2000,Inoue2005,Narayanan2018}.

In forward modelling the \flares\ galaxies, we adopted an attenuation curve which was similar to the SMC curve, but less steep, as described in Equation~\ref{eq:tau_lambda}.
In Figure~\ref{fig: att_curve}, 
we show the shape of the attenuation curve for the \flares\ galaxies at $z\in[5,10]$ in different SFR bins. 
One immediate insight is that the attenuation curve gets steeper with decreasing SFR. The galaxies with the lowest star formation rate at $z>8$ have a similar shape to our input extinction curve (Equation~\ref{eq:tau_lambda}). In case of the lower redshift galaxies, the shape becomes steeper than our input curve as well as the SMC curve. This has also been shown to be the case by other simulation works \cite[see][]{Narayanan2018} in terms of the V-band optical depth. This is mainly due to the optical light of the galaxy being dominated by old stars with less dust obscuration. The young stars in these galaxies are formed in regions of higher metallicity. Since our dust model links the metal column density to the dust optical depth\footnote{and in \cite{Narayanan2018} the radiative transfer calculations assume a dust-to-metal ratio of 0.4}, this will cause these stars to exhibit higher dust obscuration. 
This increases the A$_{\lambda}/$A$_{\mathrm{V}}$ value in the UV, thus giving rise to steeper attenuation curves. 

For highly star-forming galaxies the opposite effect is apparent, with the shape lying very close to the Calzetti starburst curve. This is not surprising considering our discussion based on our toy galaxies, as well as previous works \cite[\eg][]{Varosi1999,Witt2000,Inoue2005,Lin2021} that have already shown that the attenuation law through a clumpy medium or with increasing density along the line-of-sight with SMC dust type can look similar to the Calzetti law. As we saw in the previous section, there is a direct correlation between clumpiness and higher SFR for galaxies in \flares. The effect of this correlation can be seen from the fact that with the increasing star formation rate, the resultant attenuation curve becomes flatter. Another feature seen is the higher normalisation of the resultant attenuation curve at longer wavelengths for highly star-forming galaxies, similar to what we saw for our toy galaxies, which had higher spread in the dust attenuation. The change in the steepness of the attenuation curve between star-forming and passive galaxies was also shown in \cite{Wuyts2011}, who attributed it to changes in the dust grain size distribution, while other works \cite[\eg][]{Inoue2005,Chevallard2013} have shown that different dust optical depths can also contribute to that effect. Here we demonstrate the effect of the latter, with changes in the attenuation curve arising out of geometrical effects rather than changes in the dust grain properties.

The above results imply that by using a single extinction or effective attenuation curve for galaxies with varying SFR or UV attenuation can lead to incorrect derivation of physical properties (discussed further in \S\ref{sec: obs_cons}).

Fig.~\ref{fig: att_curve_IAB} is the same figure as \ref{fig: att_curve}, but with a selection criteria imposed to mimic the galaxy sample in \cite{Scoville2015} (galaxies brighter than I$_{\mathrm{AB}}$=25 magnitude and in the redshift range $[4.77, 6]$), to directly compare to their observed dust attenuation curve. We can see that there is a good match to the shape of the curve at wavelengths short of the rest-frame near-UV, where the curves agree very well within the spread. At longer wavelengths the normalisation is slightly higher compared to the plotted attenuation curves.  This implies that in the bright \flares\ galaxies there is either more dust attenuation at longer wavelengths, or the attenuation at 1300\AA\ is lower. This is a direct consequence of the clumpy dust geometry in the star forming galaxies in \flares, as shown also in our toy galaxies, due to the resultant A$_{\mathrm{V}}$ values moving towards lower values. We also tested using our toy galaxy sample that this persists even when excluding contribution from nebular emission.

It should be noted that the input extinction curve we used did not have the UV-absorption feature (commonly attributed to absorption by PAHs) to produce the bump feature, however, the output curve does indeed show a very weak UV-bump similar to the \cite{Scoville2015} curve. The presence of the bump feature is also visually magnified due to our normalisation being higher at longer wavelengths. Alternative explanations of the bump also include graphite grains, which are included in our nebular emission model. Even though the graphite absorption is not exactly centered at 2175\AA, the presence of graphite can thus introduce a weak UV-bump in our high-redshift sample of bright galaxies.

\section{Some observational consequences}\label{sec: obs_cons}
In this section we will discuss a few of the observational consequences of the effects described in the previous sections. 
The spread in age and chemical composition of star-forming regions, as well as the variations in their dust attenuation, have an impact on emission line ratios and the inferences drawn from analysing them.
To explore this we will concentrate on two well used line ratio methods, the Balmer decrement to correct for dust attenuation, and the BPT diagram to classify galaxies into star-forming and AGN dominated. Finally, we also look at the effect that viewing angle has on the photometry by analysing the spread in UV attenuation, UV-continuum slope and the Balmer decrement for different galaxy viewing angles.
\subsection{Balmer Decrement}\label{subsec:balmer}
\begin{figure*}
    \centering
	\includegraphics[width=0.95\textwidth]{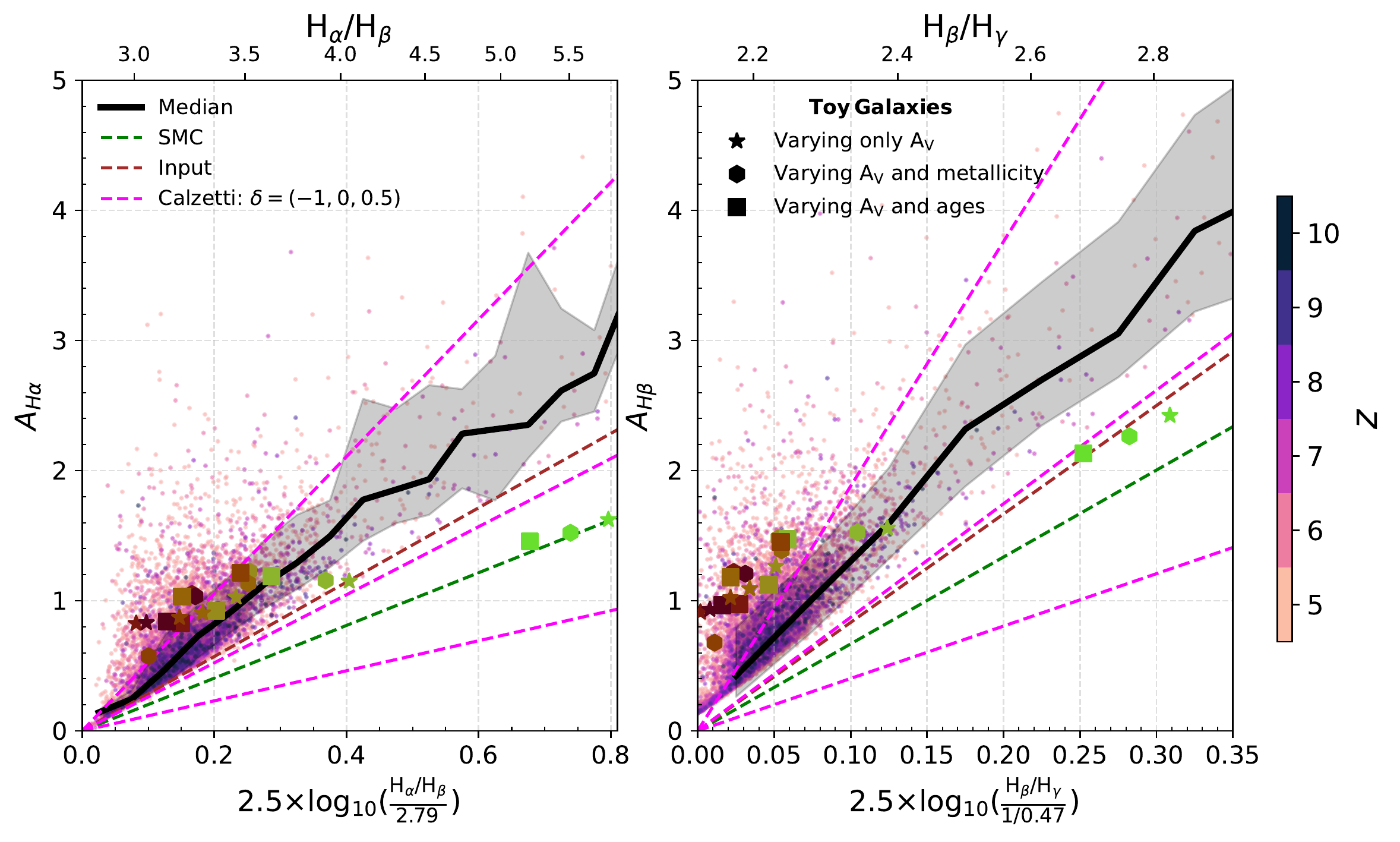}
	\caption{Balmer decrement inferred H$_{\alpha}$ (left) and H$_{\beta}$ (right) attenuation for the \flares\ galaxies in $z\in[5,10]$. The galaxies are coloured by their redshift. For reference also plotted is the median relation of the \flares\ galaxies. The expected relation for the Calzetti-starburst (for slopes of -1, 0 and 0.5), SMC and input (for the \flares\ galaxies) attenuation curves are plotted. The star, square and hexagon shaped coloured points are the relationship for the generated toy galaxies by only varying the dust attenuation towards the stars, by varying the metallicity and dust attenuation of stars, and varying the ages and dust attenuation of the stars respectively (see text for more details). \label{fig: balmer_att}}
\end{figure*}
\begin{figure*}
    \centering
	\includegraphics[width=0.95\textwidth]{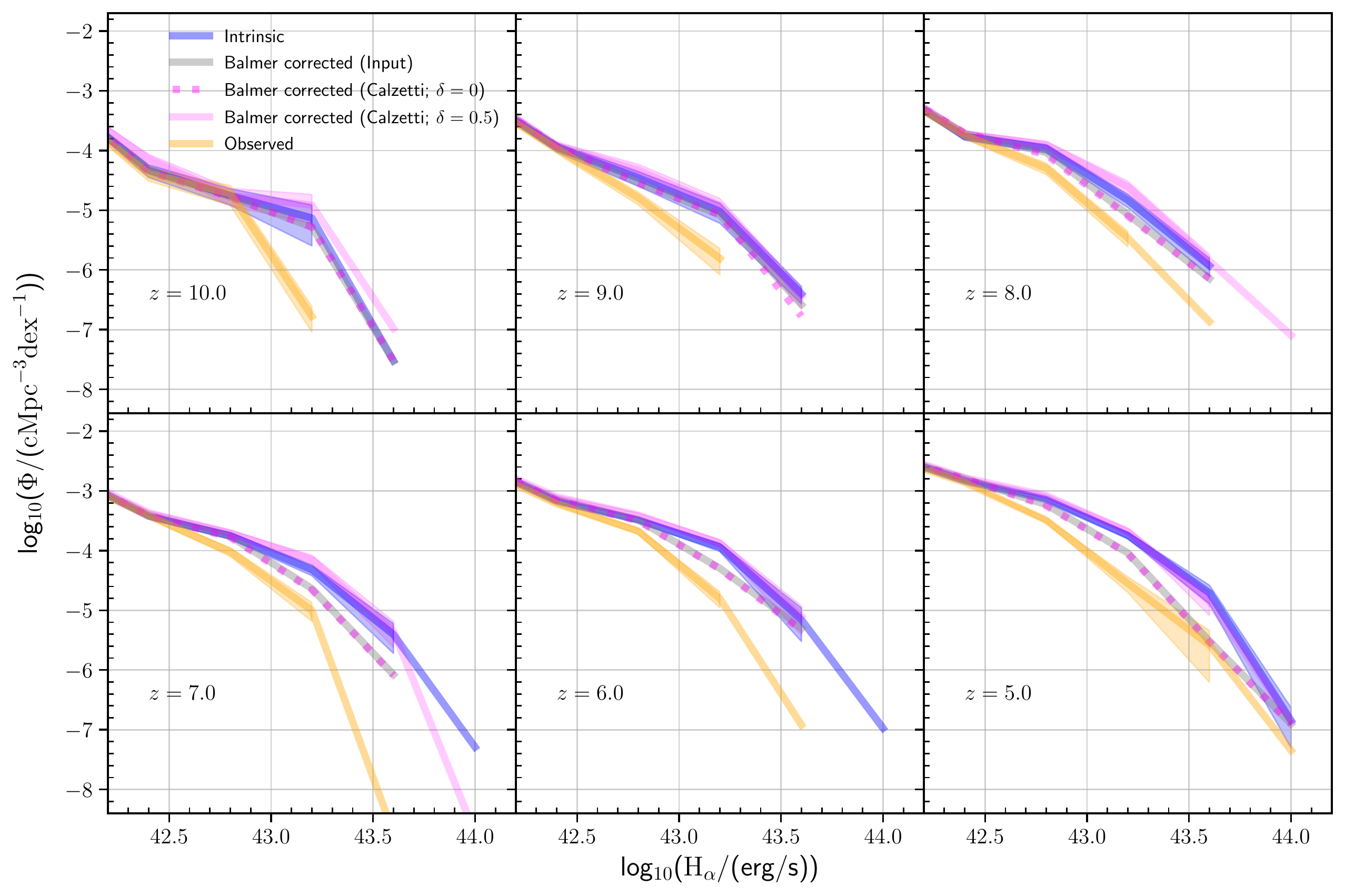}
	\caption{\Halpha\ luminosity functions of galaxies in the \flares\ for $z\in[5,10]$. The intrinsic and the dust attenuated value from the \flares\ galaxies are plotted in blue and orange colours respectively, with the shaded region denoting the Poisson 1-$\sigma$ errorbar (for only bins with more than 5 data points). Plotted alongside is the \Halpha\ luminosity function obtained from using the Balmer decrement (using the observed \Halpha\ and \Hbeta\ ratios from \flares) corrected \Halpha\ applying the \flares\ input attenuation curve and the Calzetti-starburst \protect\cite[]{Calzetti2000} curve (for $\delta=0$ and $0.5$). Note that the star particle cut mentioned in § 2.1 is not enforced here to preserve the full shape of the luminosity function. Also the Poisson 1-$\sigma$ errorbar has only been shown for the Calzetti $\delta=0.5$ curve. \label{fig: balmer_LF}}
\end{figure*}
\begin{figure*}
	\centering
	\includegraphics[width=0.95\textwidth]{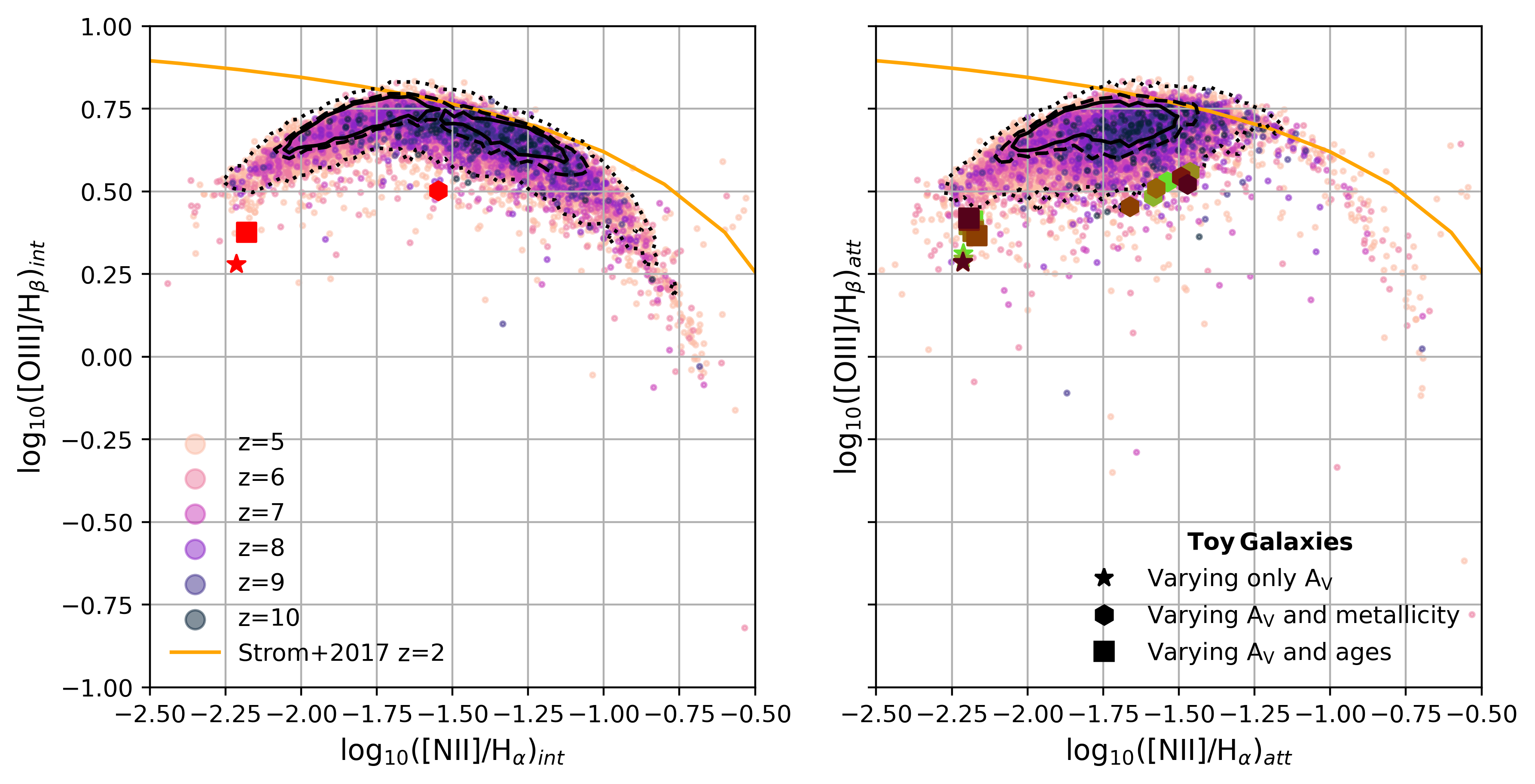}
	\caption{BPT diagram for galaxies in $z\in[5,10]$. The galaxies are coloured by their redshift. For reference also plotted is the median relation from the Keck Baryonic Structure Survey (KBSS) at $z\sim2$ \protect\cite[]{Strom2017}. The star, square and hexagon shaped coloured points are the relationship for the generated toy galaxies by only varying the dust attenuation towards the stars, by varying the metallicity and dust attenuation of stars, and varying the ages and dust attenuation of the stars respectively. \label{fig: BPT_diagram}}
\end{figure*}
The Balmer decrement, is calculated as the ratio of the Balmer transitions (transitions to the second energy level of hydrogen), usually using H$_{\alpha}$ (level $3\xrightarrow{}2$, 6562.81\AA), H$_{\beta}$ (level $4\xrightarrow{}2$, 4861.33\AA) and H$_{\gamma}$ (level $5\xrightarrow{}2$, 4340.46\AA), with each becoming progressively weaker. These transition strengths can be calculated as they are determined by atomic constants. Thus the ratio of these lines (flux or luminosity value) are usually used as a good indicator of the amount of dust attenuation within a galaxy. The dust-free or intrinsic line ratios are
\begin{equation}
    (\Halpha/\Hbeta)_{\mathrm{int}} = 2.79, 
    (\Hgamma/\Hbeta)_{\mathrm{int}} = 0.47,
\end{equation}
for electron density $10^{2} \, \mathrm{cm}^{-2}$ and temperature $15,000\,$K, conditions typically expected for high-redshift galaxies \cite[]{Sanders2023}. These ratios have a weak dependence on the electron density and are more sensitive to the temperature. However, these variations are small compared to those that can be introduced by the presence of dust \cite[]{Groves2012}.

The colour excess due the effect of dust on the observed ratios can be written as,
\begin{equation}\label{eq: balmerdecrement}
	E(\mathrm{H}_{i} - \mathrm{H}_{j}) = A(\mathrm{H}_{i}) - A(\mathrm{H}_{j}) = -2.5\,\mathrm{log}_{10}\bigg[\frac{(\mathrm{H}_{i}/\mathrm{H}_{j})_\mathrm{obs}}{(\mathrm{H}_{i}/\mathrm{H}_{j})_\mathrm{int}}\bigg],
\end{equation}
where $i$ and $j$ stands for the different Balmer lines ($\alpha$, $\beta$ and $\gamma$), with the subscripts `int' and `obs' corresponding to the intrinsic (dust-free) and observed ratios of the Balmer lines, respectively. 
Using,
\begin{equation}\label{eq: Alam}
	A(\lambda) = k(\lambda)E(B-V),
\end{equation}
where $k(\lambda)$ is the shape of the attenuation curve, one can then write,
\begin{equation}\label{eq: EBV}
	E(B-V) = \frac{2.5}{k(\mathrm{H}_{j}) - k(\mathrm{H}_{i})} \mathrm{log}_{10}\bigg[\frac{(\mathrm{H}_{i}/\mathrm{H}_{j})_\mathrm{obs}}{(\mathrm{H}_{i}/\mathrm{H}_{j})_\mathrm{int}}\bigg].
\end{equation}
By using an assumed attenuation curve, the values of $k(\mathrm{H}_{i})$ and $k(\mathrm{H}_{j})$ can be derived to infer the colour excess. 

Figure~\ref{fig: balmer_att} shows the relationship between the Balmer decrement (here quantified based on the negative of Equation~\ref{eq: balmerdecrement}) and the attenuation (calculated by integrating the dust corrected luminosity from each star particle as described in \S~\ref{sec:sim.sed}) suffered by the \Halpha\ (left) and \Hbeta\ (right) line for the \flares\ galaxies in $z\in[5,10]$. The points are coloured by their redshift. Also shown is the expected relations arising from the Calzetti and SMC attenuation curves, as well as our input curve for values of the balmer decrement using Equations~\ref{eq: Alam} and \ref{eq: EBV}. In case of the Calzetti attenuation curve, we also plot 2 modifications of it, by changing the slope of the curve by multiplying it with a wavelength dependent power law, with index $\delta$ \ie\ $\lambda^{\delta}$. Now one can modify the Calzetti curve ($\delta=0$), to a steeper ($\delta=-1$) or shallower ($\delta=0.5$) one. 

Also plotted on the figure (coloured stars) are the values for the toy galaxies presented in \S\ref{sec: attcurve} as a guide to understand the relationship between the variation in attenuation within a galaxy and the deviation from expected relation. We now add a few modifications to the intrinsic spectrum of the galaxies, by introducing a spread in the stellar metallicities (hexagons) and stellar ages (squares) of the star forming clumps. This means for one sample we vary the metallicities, keeping the ages fixed, while in the other sample, we vary the ages, keeping the metallicities fixed. The metallicities and ages are randomly sampled from a uniform distribution spanning [0.01Z$_{\odot}$, 1Z$_{\odot}$] and [1Myr, 10Myr] respectively. We then perform the same activity as in \S~\ref{sec: attcurve}; obtain the observed spectrum of the galaxies by varying the A$_{\mathrm{V}}$ magnitude of the different clumps. 

We see that \Halpha\ and \Hbeta\ attenuation is well correlated with the Balmer decrement, with Pearson correlation coefficients of $\sim0.85$. 
However, as we can see from the figure, the correlation is in-fact accompanied by large scatter, even at very low values ($\lesssim0.05$, where we expect to have very low dust attenuation), from the expected relation under assumption of any particular attenuation/extinction curve. In case of the \Hbeta/\Hgamma\ ratio, the scatter is higher at low values compared to the \Halpha/\Hbeta\ ratio.
When the Balmer decrement approaches extremely low values ($\sim 0$), the median attenuation of the lines is similar to our input curve. However, as the Balmer decrement increases, the inferred line attenuation deviates from both the expected values on the input curve as well as from all the plotted theoretical curves, trending towards higher values. It can also be seen that, even when the observed Balmer decrement is small, the discrepancy from the expected line attenuation can be a factor of $\gtrsim 10$. This is not unexpected, since the relation assumes that all the line emitting regions are equally attenuated by dust. We have already seen that differential attenuation arising from star-dust geometry, can indeed change the shape of the effective attenuation curve. This is indeed seen in Figure~\ref{fig: balmer_att}, with the shallower Calzetti curve ($\delta=0.5$) tracing the galaxies better. The steepening of the attenuation curve with increasing dust attenuation has also been shown for the high-redshift ($z\in[4.4,5.5]$) ALPINE galaxy sample in \cite{Boquien2022}. However, we note here that, in many cases the high value of $\delta=0.5$ too fails to capture the scatter fully.
Therefore, in many cases, using these line ratios can be somewhat circular, since we do not know \textit{a priori} the effective attenuation curve of the galaxy. It should also be noted that young star-forming regions (nebular emission regions) are expected to suffer higher dust attenuation than the older stars, hence the estimates from simple screen type attenuation curves can be underestimated. 

Another reason for the scatter is the distribution of stars with different stellar ages and metallicities, due to variations in star formation histories. This can also be seen from the figure, where we have plotted our toy galaxies where we varied the stellar metallicity and age, adding dust distribution variations on top of that. However, it is hard to draw conclusions on how exactly such variations skew our inferences. This has also been seen from SDSS-IV MaNGA \cite[]{Ji2023} analysis of dust attenuation from using different optical-to-NIR emission lines. They also infer that no single attenuation curve from literature can explain the attenuation of all the nebular lines within their different galaxies. \cite{Chen2023} using \jwst\ NIRCam data have already provided some insight regarding the use of line ratios to trace reionisation era galaxy properties. In their work they study resolved galaxy images, and probe the ages and optical depth of the observed clumps within the galaxy to understand the variation of the [O\textsc{iii}]$\lambda4959,5007$+H$\beta$ line strength across the galaxy, which is a probe of young star forming regions. They conclude that the variation in ages (and thus metallicities) can have a large impact on the inferred average properties of galaxies.

In Figure~\ref{fig: balmer_LF}, we plot the intrinsic and observed \Halpha\ luminosity function for \flares\ galaxies in $z\in[5,10]$. We plot alongside the Balmer decrement derived \Halpha\ luminosity function using our input attenuation curve as well as Calzetti curve with powerlaw slopes of $\delta=0$ and $0.5$, due to better match with the bulk of the data points in Figure~\ref{fig: balmer_att}. As can be seen from the figure, the input as well as the the Calzetti $\delta=0$ attenuation curve produce very identical dust-corrected luminosity functions, while the Calzetti $\delta=0.5$ curve produces a higher correction than the both of them. The effect of this can be clearly seen as we progress down in redshift. At $z=10, 9$ and $8$, all the used curves manage to reproduce the intrinsic luminosity function well, owing to the fact that there is less dust obscuration in most galaxies at these redshifts. At $z=7, 6$ and $5$, all the curves agree at the fainter end ($\le10^{42.5}$erg/s), while at the brighter end, the \flares\ input curve as well as the Calzetti $\delta=0$ curve underestimate the dust correction, while the Calzetti $\delta=0.5$ curve provides a reasonable match. This implies that given different attenuation curve choices, one can get reasonable dust corrections by employing a very shallow curve.

\subsection{BPT diagram}\label{subsec:BPT}
The BPT (Baldwin, Phillips \& Terlevich) diagram \cite[]{BPT1981}, is a set of nebular emission line ratio diagrams used to distinguish the most dominating ionising source within a galaxy. The most famous version, uses the nebular line ratios of [O\textsc{iii}]$\lambda 5008/$H$_{\beta}$ vs [N$\textsc{ii}$]$\lambda 6585/$H$_{\alpha}$ (referred to as N2-BPT) to classify galaxies into star-forming and AGN hosts. Several studies \cite[\eg][]{Brinchmann2008,Shapley2015,Strom2018,Garg2022} have looked at the loci of the galaxies inhabiting this space to understand how differences in the ionisation parameter, abundance ratios and metallicities contribute to galaxies moving across this space. Here we will look at how the variation in dust attenuation across different stars in a galaxy affect the BPT diagram. 

Figure~\ref{fig: BPT_diagram} shows the BPT diagram of the intrinsic line ratios (dust-free, left) and the dust attenuated line ratios (right). It can be seen that the locus of the points are very similar, however the density of the points has now shifted towards the left side in the dust attenuated plot compared to the intrinsic one. 
We can quantify the shift in terms of the median difference in the absolute values of the intrinsic and observed line ratios. The median shift in log$_{10}$([N$\textsc{ii}$]$\lambda 6585/$H$_{\alpha}$) is 0.08 with a standard deviation 0.2, while in the case of log$_{10}$([O\textsc{iii}]$\lambda 5008/$H$_{\beta}$) the median shift is 0.02 with a standard deviation of 0.05 across the plotted redshift range.
It is quite reasonable to wonder why there is such a difference in the intrinsic and observed line ratios, when the ratios are calculated for emission lines that are very close in wavelengths. The variation occurs because galaxies are a composite mixture of stellar populations having different ages and metallicities, which in turn determine the intrinsic line strengths of [N$\textsc{ii}$]$\lambda 6585$ and \Halpha\ or O\textsc{iii}]$\lambda 5008$ and \Hbeta. The observed (\ie attenuated) line strength depends on the dust along the line-of-sight to these different stellar populations. The total line strength is then the sum of these individual values. However, in the BPT diagram, the x and y axes represent the ratio of the sums of these individual line strengths, not an average of these individual ratios. The variation in line ratios occurs because the sum of observed ratios is not exactly equal to the sum of the intrinsic ratios multiplied by the effective attenuation \ie 
\begin{equation}\label{eq: BPTratio}
    \centering
    \frac{\sum_{k}\mathrm{F}_{i,k}\,\times\,\mathrm{T}_{i,k}}{\sum_{k}\mathrm{F}_{j,k}\,\times\,\mathrm{T}_{j,k}} \neq \frac{\sum_{k}\mathrm{F}_{i,k}}{\sum_{k}\mathrm{F}_{j,k}} \times \frac{\mathrm{T}_{\mathrm{eff},i}}{\mathrm{T}_{\mathrm{eff},j}},
\end{equation}
where F$_{i,k}$ and F$_{j,k}$ are the line fluxes of the respective lines used to compute the ratios from the k$^{\mathrm{th}}$ component in the galaxy. T$_{i,k}$ is the attenuation experienced by the line `i' for the k$^{\mathrm{th}}$ component, whereas T$_{\mathrm{eff}}$ is the assumed effective attenuation for the galaxy. 
Since the ratio $\mathrm{T}_{\mathrm{eff},i}/\mathrm{T}_{\mathrm{eff},j}$ is calculated for wavelengths that are very similar, it can be roughly approximately as 1. 
Therefore significant deviations from this assumption of effective attenuation occurs when there is a varying amount of dust along the line-of-sight of stellar populations with differing physical properties.

The seen shift in location can be explored with the toy galaxies, by plotting them on Figure~\ref{fig: BPT_diagram} (same marker types as in \S~\ref{subsec:balmer}, Figure~\ref{fig: balmer_att}). If we look at the simplest case of varying dust attenuation and keeping all the other properties same, we can see that a spread in dust attenuation moves the point vertically (up and down). Once you add to this, a spread in the metallicity of the stars, this moves the point along the right-to-left diagonal while a spread in ages moves the point along the left-to-right diagonal. Thus we can see that a combination of these effects can actually move a point from the intrinsic space towards the left in the observed space. Therefore, when interpreting physical conditions based on line ratios in dusty galaxies, it is advisable to exercise caution, even when dealing with line ratios from wavelengths that are closer to each other.

\subsection{Line-of-sight variation}\label{subsec:los var}
\begin{figure*}
	\centering
	\includegraphics[width=0.495\textwidth]{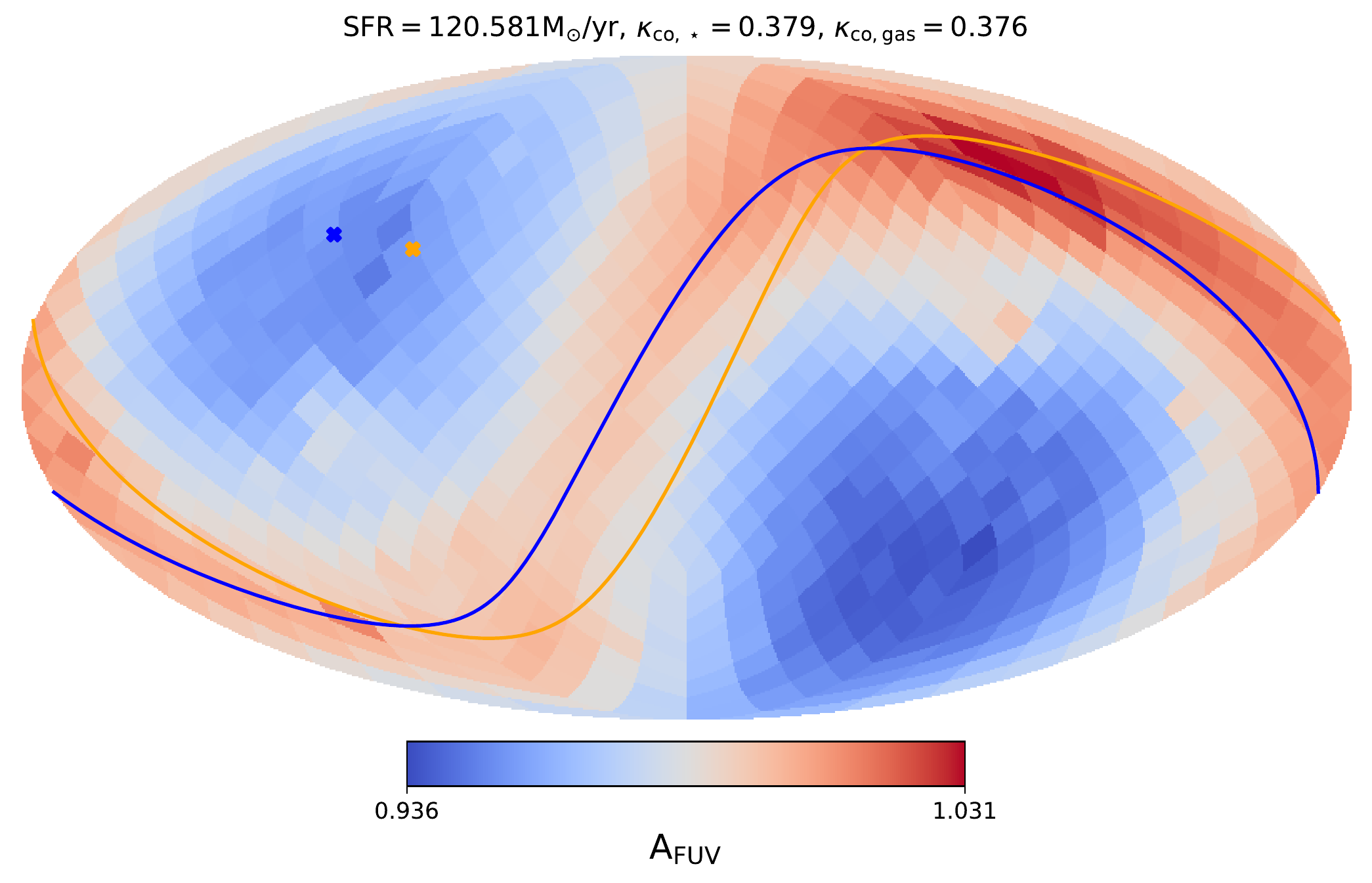}
	\includegraphics[width=0.495\textwidth]{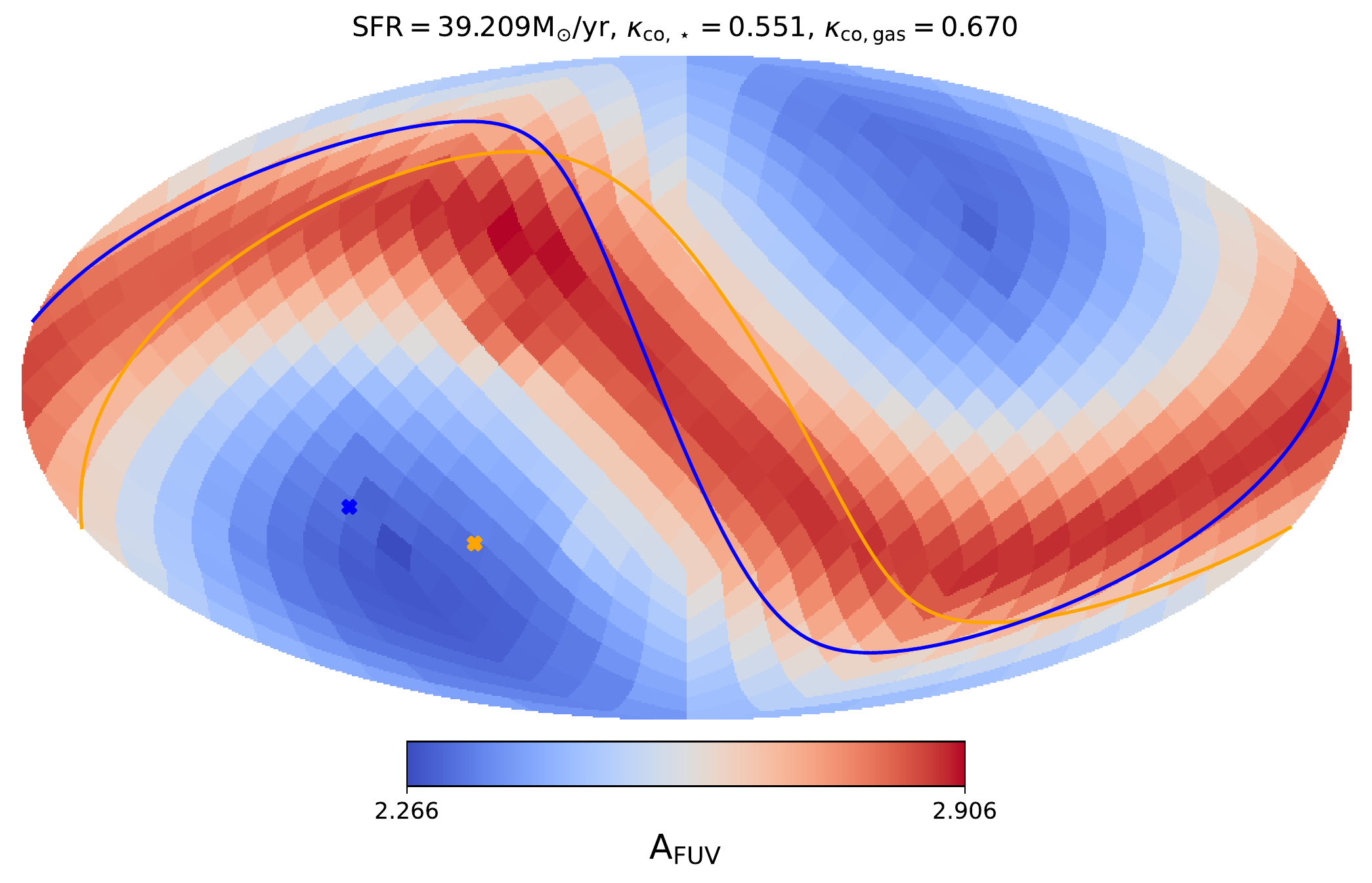}
	\includegraphics[width=0.495\textwidth]{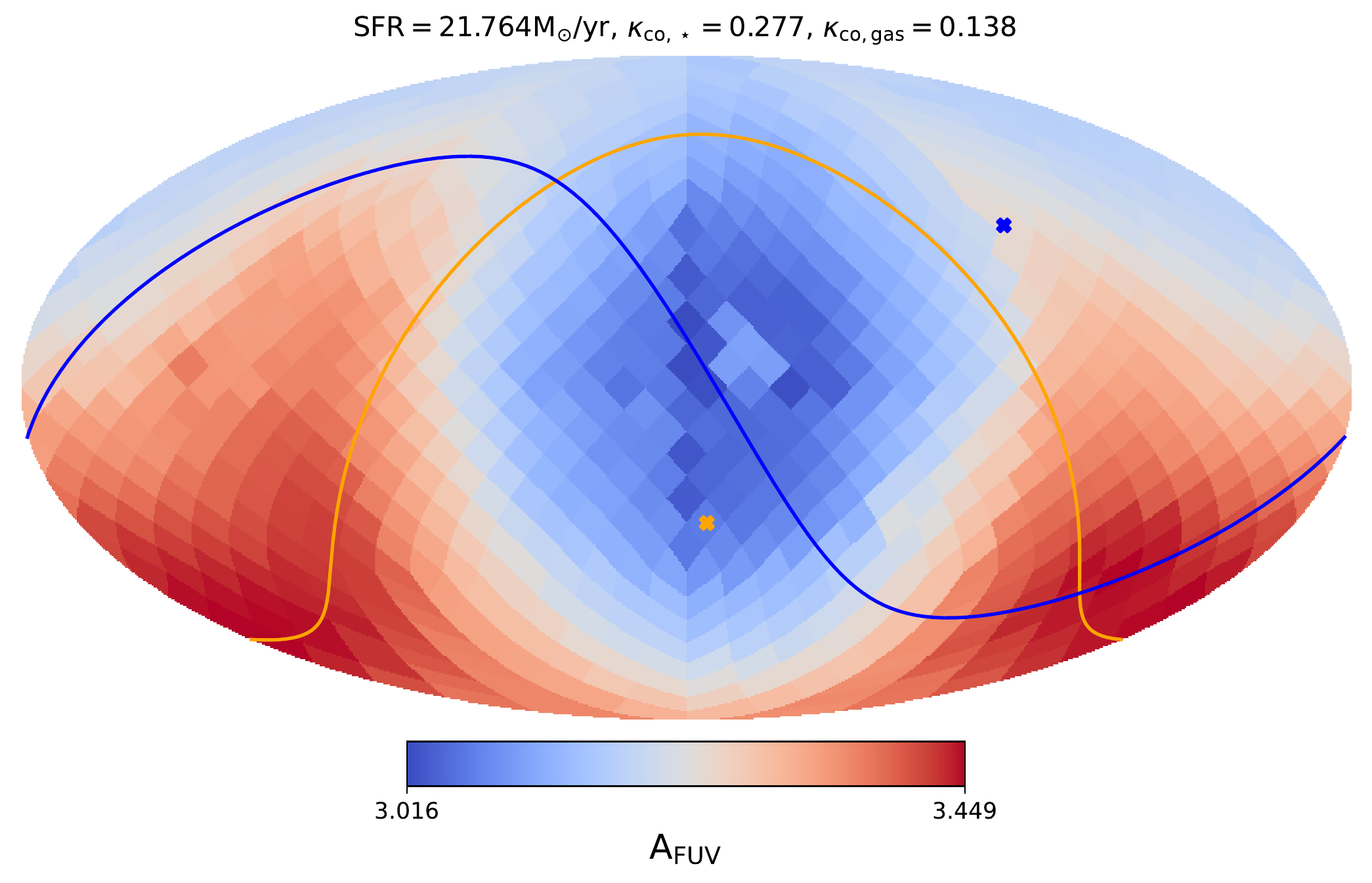}
	\includegraphics[width=0.495\textwidth]{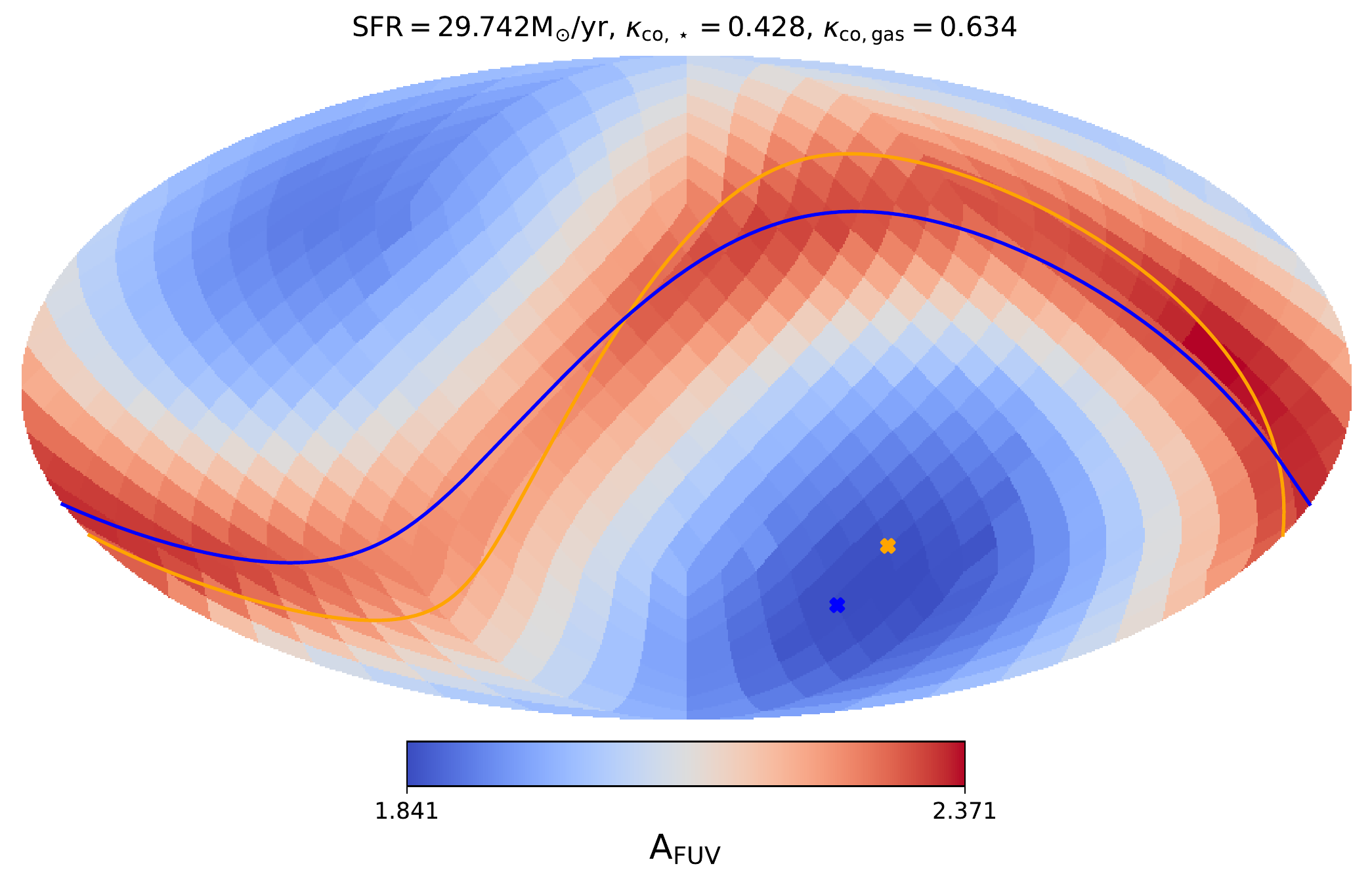}
	\includegraphics[width=0.495\textwidth]{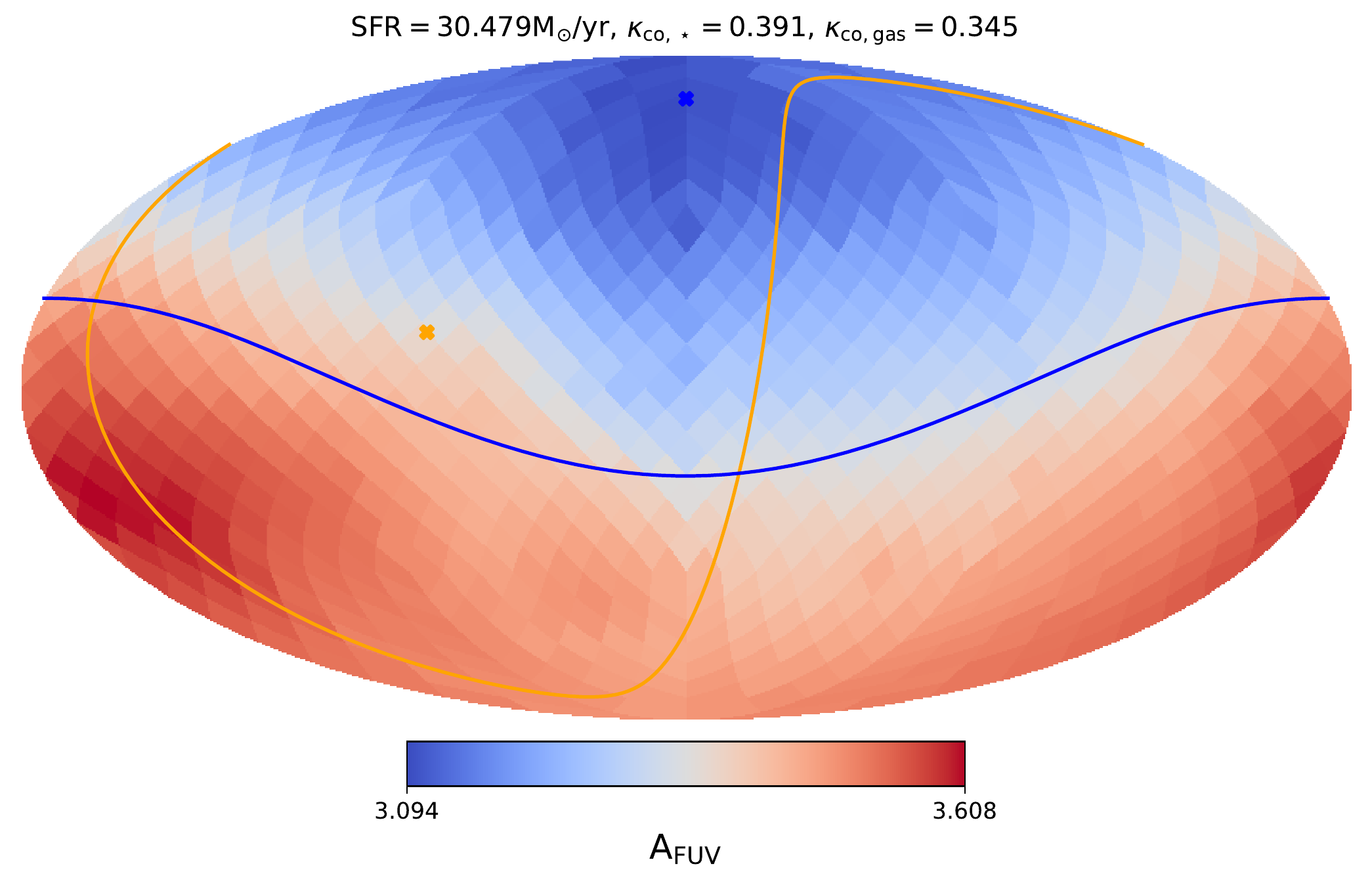}
	\includegraphics[width=0.495\textwidth]{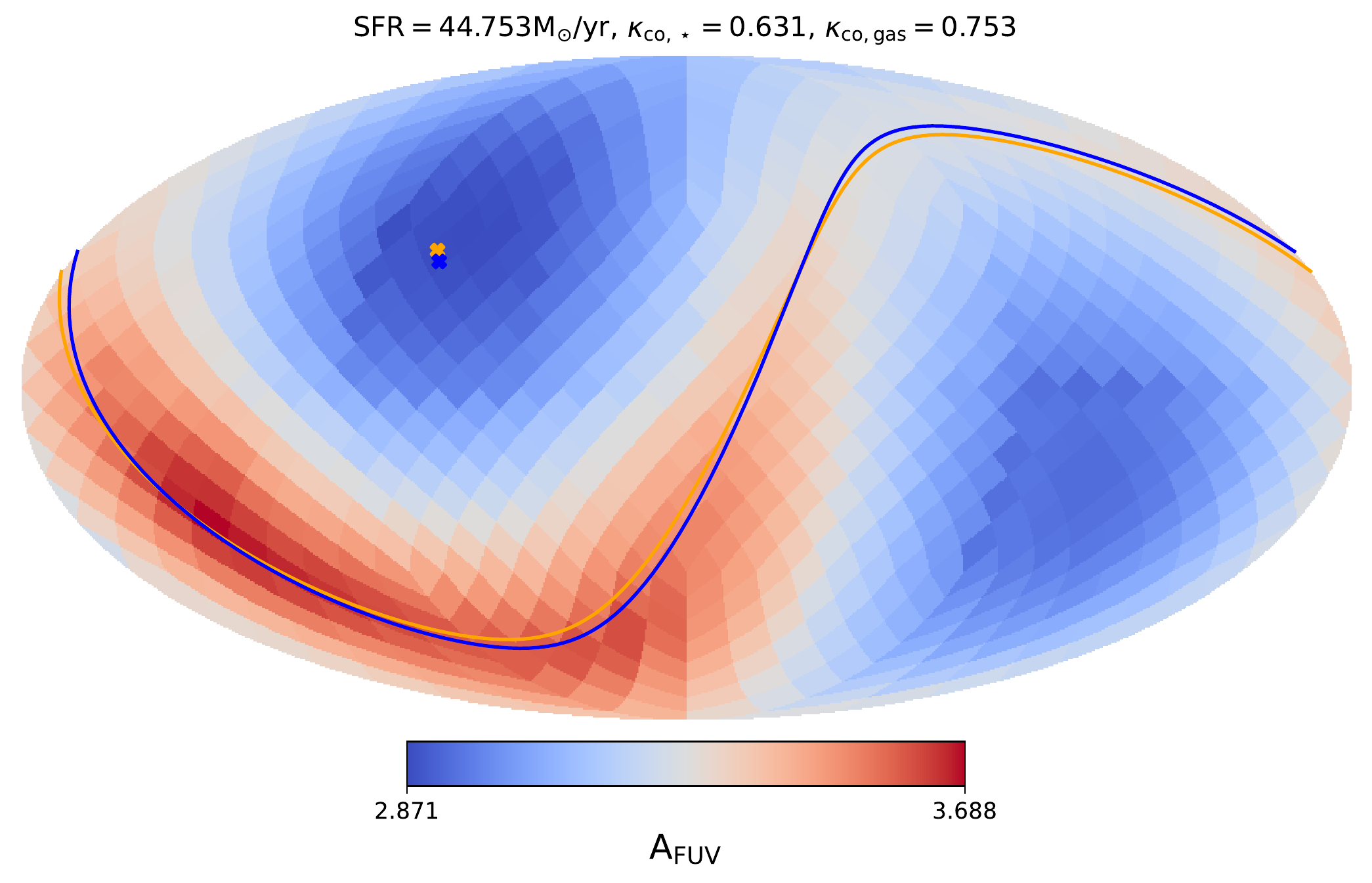}
	\caption{Healpix maps for different sightlines for $6$ galaxies at $z=5$ in our sample. The sightlines have been coloured by the galaxy UV attenuation. The blue and yellow crosses denote the sightlines looking along the angular momentum unit vector corresponding to the gas and stellar particles within the galaxy, with the lines denoting the plane perpendicular to the unit vectors. Also shown are the SFR, $\kappa_{\mathrm{co},\star}$ and $\kappa_{\mathrm{co, gas}}$ values.}
	\label{fig: healpy_plots}
\end{figure*}
The viewing angle can also change the observed properties of the galaxies that we have discussed in the previous section. This is because the dust distribution seen by spatially distinct stellar population is highly dependent on the viewing angle. This effect of the viewing angle can be explored with hydrodynamical simulations on various observed galaxy properties. In the previous sections, all the discussion assumed a single line of sight (\ie along the z-axis) to the galaxy. In this section we investigate the variation in a few galaxy observables, such as the UV-attenuation, the UV-continuum slope and the balmer decrement, when our sample galaxies are observed along different lines-of-sight (as described in \S~\ref{sec:sim.reproject}).

In $\S$\ref{sec:sim.reproject} we explained how by using \textsc{healpy} we calculate different line-of-sight re-projections. In Figure~\ref{fig: healpy_plots} we show examples of UV-attenuation maps for the different lines-of-sight for $6$ galaxies in \flares\ at $z=5$. The sightlines, represented by the healpix bins, have been coloured by the galaxy UV attenuation. The blue and yellow crosses denote the sightlines looking along the angular momentum unit vector corresponding to the gas and stellar particles, respectively, of the galaxy disc. The blue and yellow lines denote the plane perpendicular to the corresponding angular momentum unit vectors, aligning with sightlines going through the gas and stellar disc of the galaxy. 

\subsubsection{Population Trends}
\begin{figure*}
	\centering
	\includegraphics[width=\textwidth]{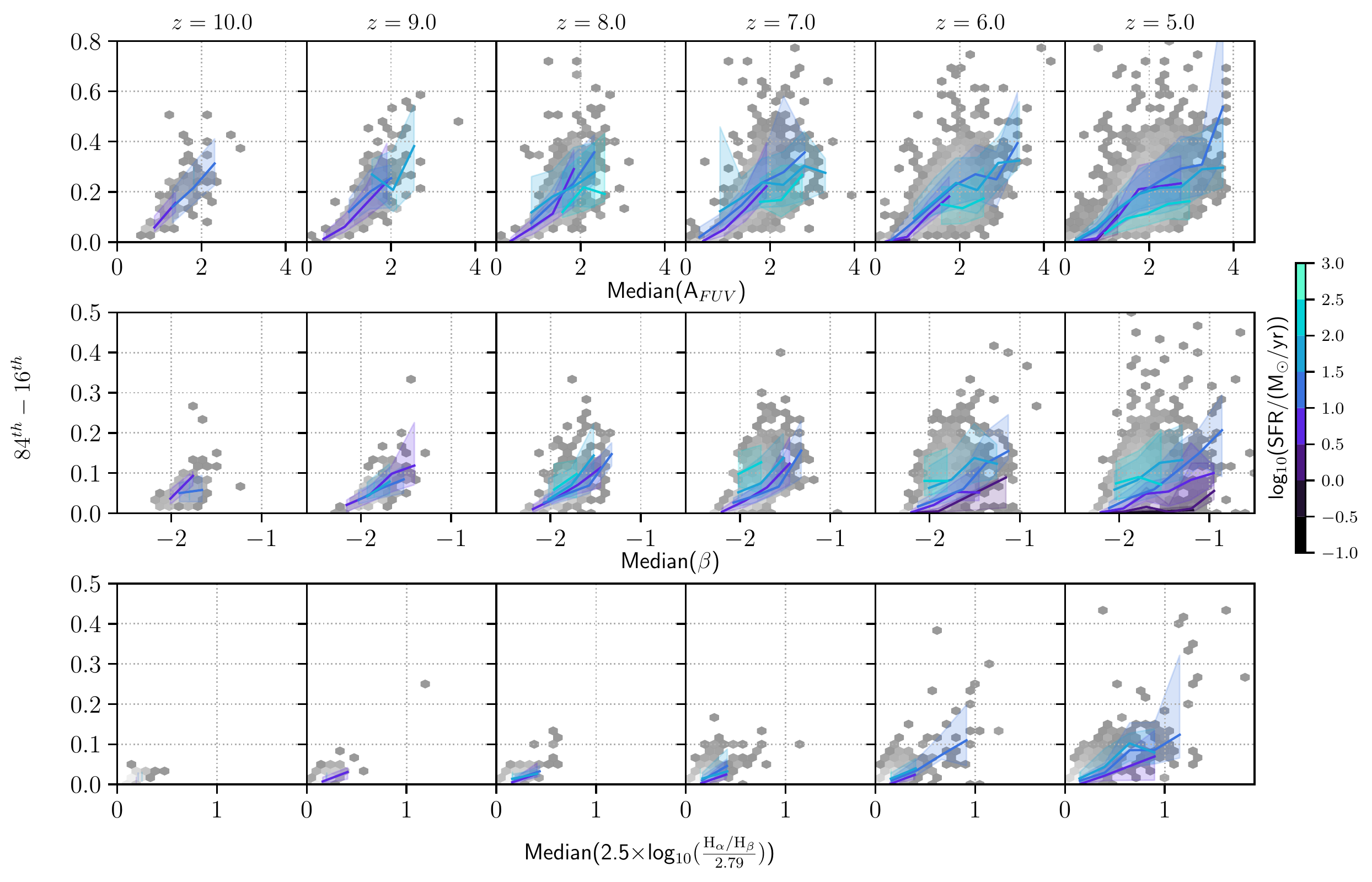}
	\caption{Shows the spread (84-16 percentile) of UV attenuation, $\beta$ and Balmer decrement values for different lines-of-sight as a function of their median values for $z\in[5,10]$. The coloured lines are the median values for different SFR$_{\mathrm{Total}}$ bins, only shown for bins with $\ge5$ galaxies. \label{fig: los_spread_SFR}}
\end{figure*}

In Figure~\ref{fig: los_spread_SFR}, we plot the $1\sigma$ spread (the difference between the $84^{th}-16^{th}$ percentile), across different lines-of-sight, of the UV-attenuation (A$_\mathrm{FUV}$), the UV-continnum slope ($\beta$) and the balmer decrement $\bigg(2.5\,\mathrm{log}_{10}\bigg[\frac{(\Halpha/\Hbeta)}{2.79}\bigg]\bigg)$ against their median values. 
The different coloured lines are the median spread for bins of different SFRs. 
From the figure it can be seen that the spread in the value of the observed A$_\mathrm{FUV}$ from its median value can be quite significant, reaching a maximum of $\sim0.4$ at $z=10$ to $\sim0.7$ at $z=5$. The variation is extremely low ($0-0.2$) at low values of the median UV attenuation ($<1$ Mag), while it can range from $0.2-0.8$ in heavily attenuated (A$_\mathrm{FUV}>2$) galaxies. It is a similar case for $\beta$ as well, reaching a maximum spread from the median $\beta$ of $\sim0.1$ at $z=10$ to $\sim0.2$ at $z=5$. Here, we also see a weak dependence of the spread on the star formation rate of the galaxy, with galaxies in higher SFR bins exhibiting a greater spread from their median value. This is another indication that the UV slope of low-SFR galaxies, or those in the passive regime, are not affected by dust to the same extent as the others.   
In case of the Balmer decrement, the variation is extremely low at $z=10$ (maximum of $\sim0.05$), with the deviation from the median value increasing with decreasing redshift (reaching a maximum of $\sim0.15$ at $z=5$). However, from Figure~\ref{fig: balmer_att} we know that there is considerable spread in the expected attenuation even at a single value of the Balmer decrement, hence small shifts in the value due to inclination effects can further add to the uncertainty of the derived parameter.
The increased dispersion around the median relation towards lower redshifts can be attributed to the build-up of dust in galaxies, which means that the star-dust geometry has a more significant impact. Additionally, there is a higher prevalence of galaxies with well-formed discs. And in such galaxies, the difference between edge-on and face-on orientations will be more pronounced.

\subsubsection{Correspondence with axis of rotation}\label{subsec: los var:discy}
\begin{figure*}
	\centering
	\includegraphics[width=\textwidth]{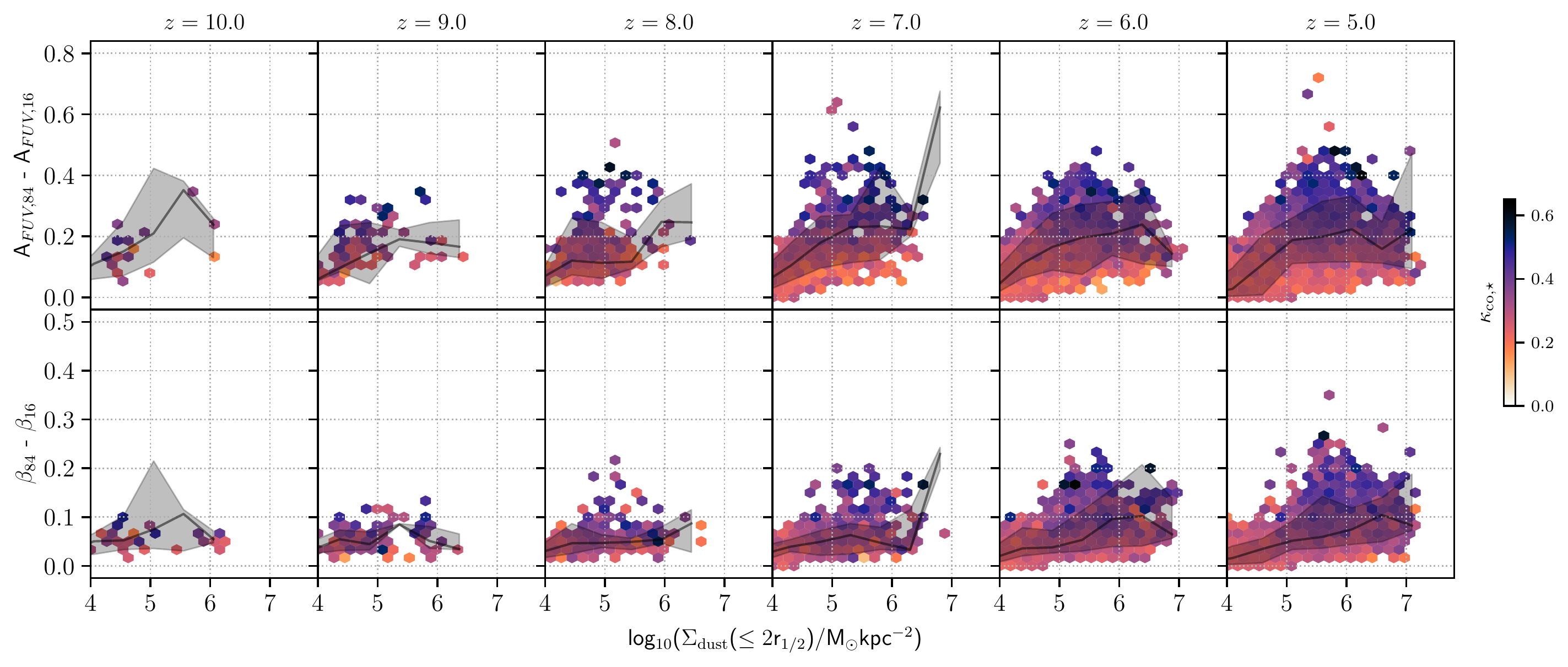}
	\caption{Shows the spread (84-16 percentile) of UV attenuation and $\beta$ values for different lines-of-sight as a function of the galaxy dust mass surface density for $z\in[5,10]$. The solid line shows the weighted median of the sample, with the shaded region denoting the 16-84 percentile, plotted for bins with $\ge5$ galaxies. \label{fig: los_spread_dsd_kappa}}
\end{figure*}
Most flagship periodic cosmological simulations (\eg \textsc{Eagle}, \textsc{Illustris-tng}, \textsc{Simba}, \etc), at the redshifts we are looking at have a dearth of disc galaxies due to the smaller physical volume they probe. Therefore, they sample fewer instances of the highly overdense regions, which are rare and expected to host the massive galaxies in the early Universe, including the formation of disc galaxies. This is not a problem in \flares, due to the novel re-simulation strategy employed, with regions ranging from extremely underdense to extremely overdense regions. And with the resimulated regions in \flares\ picked from a (3.2cGpc)$^3$ \cite[see Figure~1 in][]{Lovell2021} dark matter only box, has access to a wider range overdenisties compared to typical flagship periodic cosmological simulations.
There are several methods available for identifying disc galaxies in simulations \cite[for examples see][]{Irodotou2021}. However, for this work, we quantify a galaxies' disciness by computing the energy invested in ordered rotation \cite[see][for more details]{Sales2012,Correa2017}. 
% \peter{Shame on you; Dimi's method is better.} 
The following equation is used to compute the value:
\begin{equation}
	\kappa_{\mathrm{co}} = \frac{K_{\mathrm{co}}}{K} = \frac{1}{K}\,\sum_{i}\frac{1}{2}\,m_{i}\,[L_{z,i}/(m_{i}R_{i})]^2,
\end{equation} 
where the sum is over all the particles of a particular type (gas or stars) within our aperture centred on the potential minimum, $m_i$ is the mass of the particle and $K$ is the total kinetic energy ($\sum_{i}\frac{1}{2}\,m_{i}v_{i})^2$, with $v_{i}$ the velocity of the particle). $L_{z,i}$ is the particle angular momentum along the direction of the total angular momentum of the component type of the galaxy and $R_i$ the projected distance to the axis of rotation. We compute $\kappa_{\mathrm{co}}$ for the gaseous and stellar component of the galaxy, denoting them as $\kappa_{\mathrm{co, gas}}$ and $\kappa_{\mathrm{co}, \star}$ respectively.

To understand the relationship between the spread in the UV attenuation and $\beta$, we plot a hexbin distribution of these values against the dust surface density in Figure~\ref{fig: los_spread_dsd_kappa}; the dust surface is calculated within $2\times R_{1/2,\mathrm{dust}}$, where $R_{1/2,\mathrm{dust}}$ is the half-mass dust radii. 
We colour the hexbin distribution by the median $\kappa_{\mathrm{co},\star}$ value within the bin. It can be clearly seen that there is a weak evolution in the spread of both the UV attenuation and $\beta$ with the dust surface density, with the extreme variation in the spread taken up by galaxies that are very discy. There is a clear gradient in $\kappa_{\mathrm{co},\star}$ values moving from low to high spread. 
This can also be seen in Figure~\ref{fig: healpy_plots} where the maximum of the attenuation traces the line-of-sight corresponding to the plane of the disc in these galaxies \ie\ viewing through the galaxy gaseous/stellar disc.

\section{Conclusions}\label{sec:conc}
In this work we have explored how variations in dust attenuation across a galaxy, affect their observed properties using the First Light And Reionisation Epoch Simulations (\flares). Our conclusions are as follows:
\begin{itemize}
    \item We explore the radial profile of the UV traced star formation rate surface density as well as the UV attenuation, finding that the star formation in the brightest galaxies in \flares\ have extended towards the outskirts. However these stars experience less dust attenuation than the stars formed in the galaxy cores. This results in a non linear relationship between the intrinsic UV luminosity and the UV attenuation at the bright end, with the \flares\ UVLF exhibiting a double power-law shape.
    \item We also look at the spread in the UV attenuation of different star particles within a galaxy, finding huge spread, with many galaxies having stars that are fully obscured to fully unobscured. We found that the magnitude of this spread is correlated with the star formation rate of galaxies in \flares, with the extreme star forming galaxies exhibiting the largest spread, while less star-forming galaxies have more homogeneous dust distribution or very low dust attenuation.
    \item We explore the consequences of this spread by looking at the resultant attenuation curve of galaxies in \flares. We find that low SFR galaxies exhibit very steep attenuation curve while galaxies with extreme SFRs show flatter attenuation curves. 
    \item We explore some of the observational consequences of this spread in dust attenuation across the galaxy, by probing its effects on the Balmer decrement and the BPT diagram. Dust complicates the interpretation of these line ratios. The expected attenuation calculated from the Balmer decrement does not follow the widely used attenuation/extinction curves from literature. The location of points in the BPT diagram compared to the points before dust attenuation are also changed. Inhomogeneous dust distribution coupled with the variation in stellar ages and metallicities across the galaxy is the main driver of these changes.
    \item We also explore the effect of variations in the viewing angle towards the galaxy. Different viewing angles results in different star-dust geometry, which also introduces spread in the observed values of the UV attenuation, UV-slope and the Balmer decrement. The extreme variations in this space are observed for galaxies that are very discy.
\end{itemize}

With recent influx of data from \jwst\ of the high-redshift Universe, it is now possible to explore galaxy ISM conditions using nebular emission lines \cite[\eg][]{Arellano2022,Endsley2023,Cameron2023,Sanders2023}. These have already provided some insights into the ISM conditions such as the ionisation parameter, metallicity and elemental abundances. There is promise of even more spectroscopic data, with programmes covering large areas of the sky (\eg\ Cosmos-Web \cite[]{Casey2023}, UNCOVER \cite[]{Bezanson2022}), that will be able to provide statistically significant samples of galaxies. These will enable to understand the diverse conditions within various galaxy populations across redshift. However, our results suggest that we should be careful in the interpretation of these results. Galaxies are complex structures consisting of distinct regions with varying physical conditions and properties. Consequently, many commonly used diagnostic measures, such as the Balmer decrement and BPT diagram, are a composite measure of these diverse physical conditions. This diversity in properties can conspire together to yield diagnostic measurements that differ from the actual conditions present within the galaxy, ultimately impacting our conclusions.

\section*{Acknowledgements}
The authors thank the referee for their comments in improving this manuscript.
The authors wish to thank Adam Carnall and Victoria Strait for helpful discussions.  This work used the DiRAC@Durham facility managed by the Institute for Computational Cosmology on behalf of the STFC DiRAC HPC Facility (www.dirac.ac.uk). The equipment was funded by BEIS capital funding via STFC capital grants ST/K00042X/1, ST/P002293/1, ST/R002371/1 and ST/S002502/1, Durham University and STFC operations grant ST/R000832/1. DiRAC is part of the National e-Infrastructure. 

APV \& TRG acknowledges support from the Carlsberg Foundation (grant no CF20-0534). The Cosmic Dawn Center (DAWN) is funded by the Danish National Research Foundation under grant DNRF140. CCL acknowledges support from a Dennis Sciama fellowship funded by the University of Portsmouth for the Institute of Cosmology and Gravitation. DI acknowledges support by the European Research Council via ERC Consolidator Grant KETJU (no. 818930). 

We also wish to acknowledge the following open source software packages used in the analysis: \texttt{astropy} \citep{Astropy2013,Astropy2018,Astropy2022}, \texttt{cmasher} \cite[]{cmasher2020}, \texttt{matplotlib} \citep{Hunter:2007}, \texttt{numpy} \citep{Harris2020_numpy}, \texttt{pandas} \cite[]{reback2020pandas}, \texttt{scipy} \citep{2020SciPy-NMeth}. Some of the results in this paper have been derived using the \texttt{HEALPix} \cite[]{HEALPix} package.

We list here the roles and contributions of the authors according to the Contributor Roles Taxonomy (CRediT)\footnote{\href{https://credit.niso.org/}{https://credit.niso.org/}}. \textbf{Aswin P. Vijayan}: Conceptualization, Data curation, Methodology, Investigation, Formal Analysis, Visualization, Writing - original draft. \textbf{Peter A. Thomas, Stephen M. Wilkins}:  Conceptualization, Data curation, Writing - review \& editing. \textbf{Christopher C. Lovell, Dimitrios Irodotou, William J. Roper, Louise T. C. Seeyave}: Data curation, Writing - review \& editing. \textbf{Thomas R. Greve}: Methodology, Writing - review \& editing. 

\section*{Data Availability Statement}
The data underlying this article (stellar masses, star formation rates, photometry and line luminosities for $z \in [5, 10]$) are available at \href{https://flaresimulations.github.io/data.html}{flaresimulations.github.io/data}. Additional data is available upon reasonable request to the corresponding author. The code to reproduce the plots in the paper can be found at: \href{https://github.com/flaresimulations}{https://github.com/flaresimulations}.



\bibliographystyle{mnras}
\bibliography{ref.bib} % if your bibtex file is called example.bib
%%%%%%%%%%%%%%%%%%%%%%%%%%%%%%%%%%%%%%%%%%%%%%%%%%

%%%%%%%%%%%%%%%%% APPENDICES %%%%%%%%%%%%%%%%%%%%%

\appendix

%%%%%%%%%%%%%%%%%%%%%%%%%%%%%%%%%%%%%%%%%%%%%%%%%%

%%%%%%%%%%%%%%%%%%%%%%%%%%%%%%%%%%%%%%%%%%%%%%%%%%

% Don't change these lines
\bsp	% typesetting comment
\label{lastpage}
\end{document}